\documentstyle[a4,aps,epsfig]{revtex}
\catcode`\@=11
\title{\LARGE\textbf{ 
HIGHER CONFORMAL MULTIFRACTALITY}\footnote{ 
December 2001 Rutgers meeting in celebration of Michael E. Fisher's $70^{\rm th}$ birthday, 
to appear in {\it J. Stat. Phys.}}}
\author{Bertrand {\sc Duplantier}}
\address{Service de Physique Th\'{e}orique\footnote{URA CNRS 2306} \\
CEA/Saclay\\ 
F-91191 Gif-sur-Yvette Cedex\\
FRANCE}
\date{29 April 2002}
\begin{document}
\input epsf
\draft
\maketitle
\vskip1cm
\begin{abstract}
{\bf Abstract.}
We derive, from conformal invariance and quantum gravity, the multifractal spectrum $f(\alpha)$ of
the harmonic measure  (or
electrostatic potential, or diffusion 
field) near any conformally invariant fractal in two dimensions. It gives the Hausdorff
dimension of the set of points where the potential varies with distance $r$
to the fractal frontier as $r^{\alpha}$. First examples are a random walk,
i.e., a Brownian motion, a
self-avoiding walk, or a critical percolation
cluster. The
generalized dimensions $D\left( n\right) $ as well as the multifractal functions
 $f\left( \alpha 
\right) $ are derived, and are all identical for these three cases.
The external frontiers of a Brownian motion
and of a percolation cluster are thus {\it identical to a self-avoiding walk} in the
 scaling limit.
The multifractal (MF) function $f(\alpha,\,c)$ of the electrostatic potential near any
 conformally invariant fractal boundary, like a critical
$O(N)$ loop or a $Q$ -state Potts
cluster, is solved as a function of the central charge $c$
of the associated conformal field theory.
The dimensions $D_{\rm EP}$ 
of the external
perimeter and $D_{\rm H}$ of the hull of a critical scaling curve or cluster obey
the {\it superuniversal} duality equation $(D_{\rm EP}-1)(D_{\rm  H}-1)=\frac{1}{4}$.
Finally, for a conformally invariant scaling curve which is {\it simple}, i.e., without double points,
 we derive higher multifractal 
functions,
like the universal function $f_2(\alpha,\alpha')$ which gives the Hausdorff dimension of the points where the potential
jointly varies with distance $r$ as $r^{\alpha}$ on one side of the curve, and as $r^{\alpha'}$ on the other.
The general case of the potential distribution between the branches of a star made of an arbitrary
number of scaling paths is also treated. The results apply to critical $O(N)$ loops, Potts clusters, and to the
$SLE_{\kappa}$ process. We present a duality between external perimeters of Potts clusters and $O(N)$ loops 
at their critical point, 
as well as the corresponding duality in the $SLE_{\kappa}$ process for $\kappa\kappa'=16$.   
\end{abstract}

\bigskip

{\textbf{\large\textbf{I Introduction}}}\\

Quantum field theory can be
described very generally in terms of the statistics of Brownian paths and of
their
intersections \cite{symanzyk}. This equivalence is used in polymer theory
\cite{symanzyk} and in rigorous studies of second-order phase transitions and
field theories \cite{aizenman1}. In probability theory, non trivial properties of
Brownian paths have led to intriguing conjectures.
Mandelbrot \cite{mandelbrot} suggested for instance that in two
dimensions, the external frontier of a planar Brownian path has a
Hausdorff dimension $D=4/3$, identical to that of a planar self-avoiding walk,
i.e., a polymer. Families
of universal critical exponents are associated with {\it intersection} properties
of
sets of random walks\cite{fisher,lawler1,aizenman2,duplantier1,duplantier2,sokal,burdzy}. In his famous article
 ``{\it Walks, walls, wetting and melting}'' \cite{fisher}, Michael Fisher considered groups of
 1D ``vicious walkers'' which had to avoid each other: this amounts to considering 2D {\it directed}
 mutually-avoiding random walks. The present paper deals with the scaling properties
 of 2D mutually-avoiding random walks or fractal sets. Here the latter are conformally invariant,
 allowing the use of techniques borrowed from ``2D quantum gravity'', where mutual avoidance is equivalent to
 solving a linear problem. In \cite{fisher}, the directness of the walks made the
 problem tractable. The general similarity in the scaling concepts used, however, certainly allows one
 to dedicate this work as a tribute to the outstanding influence Michael Fisher had in the general
 field of random scaling paths.

The concepts of generalized dimensions and associated
multifractal (Mf)
measures have been developed two decades ago \cite
{mandelbrot2,hentschel,frisch,halsey}. Universal geometrical fractals, e.g.,
random walks,
polymers, Ising or percolation models are essentially related to standard
critical phenomena and field theory, for which conformal invariance in two
dimensions (2D) has brought a wealth of exact results
(see, e.g.,\cite{belavin,friedan,dennijs,nien,duplantier4,DS2,cardy3}. Multifractals and field theory must have 
deep connections, since the algebras of their respective correlation functions reveal intriguing
similarities \cite{cates}. 

In classical potential theory, i.e., that of the electrostatic or diffusion
field near random fractal boundaries such as diffusion limited aggregates (DLA), or the
fractal structures arising in critical phenomena, the self-similarity of the latter is reflected in a
{\it multifractal} (MF) behavior
of the potential. In DLA, the potential,
also called harmonic
measure, actually determines the growth process and its scaling
properties are intimately related to those of the
of the
cluster itself\cite{meak}. In statistical fractals, the Laplacian field is created by the random boundary, and
should be derivable, in a probabilistic sense, from the knowledge of the latter. A first example was
studied in Ref. \cite{cates et witten}, where the fractal boundary, the ``absorber'', was
chosen to be a simple random walk (RW), or a self-avoiding walk (SAW), accessible to
renormalization group methods near four dimensions.

In {\it two dimensions} (2D), conformal field theory (CFT) has lent strong support to the conjecture
that statistical systems at their critical point, in their
scaling (continuum) limit,
produce {\it conformally invariant} (CI) fractal structures, examples of which are
the continuum scaling limits of RW's, SAW's, critical Ising or Potts clusters (which presented a
mathematical challenge, see, e.g.,\cite{langlands,ai1,ben}). The harmonic
measure near a such clusters possesses universal multifractal exponents, as we shall see.
In analogy to the
beautiful simplicity of the
classical method of conformal transforms to solve 2D
electrostatics of {\it Euclidean} domains, a {\it universal}
solution is indeed possible for the planar potential near a CI fractal. 

A first exact example has been
given for the whole
universality class of random or self-avoiding walks, and percolation clusters, which all possess
 the same harmonic MF spectrum in two dimensions \cite{duplantier7,duplantier8,duplantier9} (Related results can be found in
 \cite{lawler2,lawler3,cardy2}.)
 After a detailed description of this class and its link to quantum gravity, we address the general
 solution for the potential distribution near any conformal
 fractal in 2D \cite{duplantier11}.
 The exact multifractal spectra describing the singularities of the potential, or, equivalently,
 the distribution of local ``electrostatic'' wedge angles along the fractal boundary, are given, and shown
 to depend only on
 the so-called
 {\it central charge c}, a parameter which labels the universality class of the underlying CFT. A further result, first obtained in 
 \cite{duplantier11}, is the existence of a {\it duality} between the external frontiers of random clusters and their hulls, 
 which applies in particular to Fortuin-Kasteleyn clusters in the Potts model, and to the so-called 
 stochastic L\"owner evolution ($SLE$) process (see below).

 A new feature will be the consideration of {\it higher multifractality}, which occurs in a
 natural way in the joint distribution
 of potential on both sides of a random CI scaling path, or more generally, in the distribution of
 potential between the
 branches of a {\it star} made of an arbitrary number of CI paths. The associated universal multifractal spectrum
 will depend on two variables, or more generally, on $m$ variables
 in the case of an $m$-arm star. We shall derive it first for Brownian motion 
 or self-avoiding walks \cite{duplantier10}, before addressing the general case.

Consider a two-dimensional very large ``absorber'' ${\cal S}$, which can be a 
random walk, a self-avoiding walk, a percolation cluster, or, more generally, a (critical) scaling path or cluster. Define
the harmonic measure ${\rm H}\left( w\right) $ as the probability that a random walker (RW) launched
from 
infinity, {\it first} hits the outer ``hull's frontier'' or (accessible) frontier ${\cal F}({\cal S})$ at point $w \in {\cal F}({\cal S})$. One then considers a covering
of $\cal F$ by balls ${\cal B}(w, a)$ of radius $a$, and centered at points $w$
 forming a discrete subset ${\cal F}/\{a\}$ of $\cal F$. Let ${\rm H}({\cal F} \cap {\cal B}(w, a))$ be the
harmonic measure of the points of ${\cal F}$ in the ball ${\cal B}(w, a)$. We are 
interested in the moments of $H$, averaged over all
realizations of RW's and ${\cal S}$ 
\begin{equation}
{\cal Z}_{n}=\left\langle \sum\limits_{w\in {\cal F}/\{a\}}{\rm H}^{n}\left({\cal F}\cap{\cal B}(w, 
a)\right)
\right\rangle ,  \label{Z'}
\end{equation}
 where $n$ can be, {\it a priori},
a real number. For very large absorbers ${\cal S}$ and hull's frontiers ${\cal F}\left( 
{\cal 
S}\right) 
$ of average size $R,$ one expects these moments to scale as 
\begin{equation}
{\cal Z}_{n}\approx \left( a/R\right) ^{\tau \left( n\right) },  \label{Z2'}
\end{equation}
where the radius $a$ serves as a microscopic cut-off, reminiscent of
the lattice structure,
and where the multifractal scaling exponents 
$\tau \left(
n\right) $ encode {\it generalized dimensions}
\begin{equation}
D\left( n\right) 
=\frac{\tau\left( n\right)}{n-1} ,
\label{dn'}
\end{equation}
which vary in a non-linear way with $n$\cite{mandelbrot2,hentschel,frisch,halsey}. 
Several 
{\it 
a priori} results are known. $D(0)$ is the Hausdorff dimension of the accessible frontier of the fractal.
By construction, H is a normalized probability measure, so
that $\tau (1)=0.$ Makarov's theorem \cite{makarov}, here applied to the
H\"{o}lder regular curve describing the frontier \cite{ai2}, gives the {\it non 
trivial} information dimension $\tau ^{\prime }\left( 1\right) =D\left( 1\right) =1.$
The multifractal formalism \cite{mandelbrot2,hentschel,frisch,halsey} further 
involves
characterizing subsets ${\cal F}_{\alpha }$ of sites of the hull's frontier ${\cal F}$
by a H\"{o}lder exponent $\alpha ,$ such that the ${\rm H}$-measure of the frontier points 
in the ball ${\cal B}(w, a)$ of radius $a$ centered at $w_{\alpha}\in {\cal F}_{\alpha }$ scales as 
\begin{equation}
{\rm H}\left({\cal F} \cap {\cal B}(w_{\alpha}, a) \right) \approx \left( a/R\right) 
^{\alpha }.
\label{ha'}
\end{equation}
The Hausdorff or ``fractal dimension'' $f\left( \alpha \right) $ of the
set ${\cal F}_{\alpha }$, such that
\begin{equation}
{\rm Card}\, {\cal F}_{\alpha} \approx R^{f(\alpha)},
\label{ca'}
\end{equation}
is given by the symmetric Legendre transform of $%
\tau \left( n\right) :$%
\begin{equation}
\alpha =\frac{d\tau }{dn}\left( n\right) ,\quad \tau \left( n\right)
+f\left( \alpha \right) =\alpha n,\quad n=\frac{df}{d\alpha }\left( \alpha
\right) .  \label{alpha}
\end{equation}
Because of the statistical ensemble average (\ref{Z'}), values of $%
f\left( \alpha \right) $ can become negative for some domains of $\alpha $ 
\cite{cates et witten}. As we shall see, the associated exponents $\tau (n)$ above can be
recast as those of star copolymers made of independent RW's in a bunch, diffusing 
away from a generic point of the absorber.

One can equivalently consider potential theory near the same fractal boundary, now charged. One assumes the 
absorber to be perfectly conducting, and introduces the harmonic potential $H(z)$ at points $z$ in the domain exterior 
to $\cal F$, with (Dirichlet) boundary condition 
$H=0$ on ${\cal F}$, and $H=1$ on a large exterior circle. Then the local behavior of the potential
\begin{equation}
H (z \to w_{\alpha}) \sim r^{\alpha}, \ r=|z-w_{\alpha}|\ ,
\label{Hal}
\end{equation} 
depends on the same $\alpha$-exponent as the harmonic measure
around point $w_{\alpha} \in \cal F_{\alpha},$ and $f(\alpha)={\rm dim} \cal F_{\alpha}$ appears as
the Hausdorff dimension of boundary points inducing the local behavior (\ref{Hal}).  

\begin{figure}
\centerline{\epsfig{file=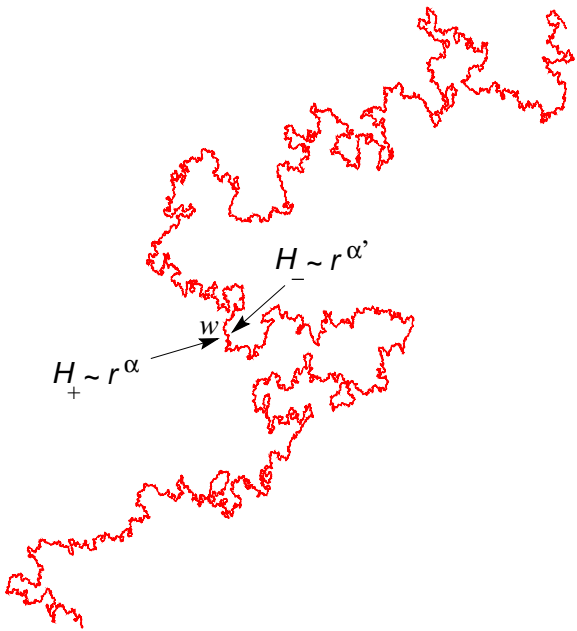,width=13cm}}
\smallskip
\caption{Double distribution of harmonic potential $H$ on both sides of a simple scaling curve 
(here a SAW, courtesy of Tom G. Kennedy). The local exponents on both sides of point $w=w_{\alpha,\alpha'}$ are $\alpha$ and $\alpha'$. The Hausdorff
 dimension of such points along the SAW is $f_2(\alpha,\alpha')$.}
\label{Figure0}
\end{figure}
Generalizations to higher conformal multifractality can be defined as follows. 
When it is {\it simple}, i.e., without double points, the conformally scaling curve $\cal F$ can be 
reached from both sides, with a distribution of potential $H_{+}$ on one side, and $H_{-}$ on the other, such that
\begin{eqnarray}
 H_{+}(z\to w_{\alpha,\alpha'}) \sim r^{\alpha}, \ H_{-} (z\to w_{\alpha,\alpha'})\sim r^{\alpha'}\ ,
\end{eqnarray}
 when approaching a point $w_{\alpha,\alpha'}$ of subset $\cal F_{\alpha,\alpha'}$  at
 distance $r=|z-w_{\alpha,\alpha'}|$ (Fig. 1).
 Then a double-multifractal spectrum $f_2(\alpha, \alpha')={\rm dim}\,\cal F_{\alpha,\alpha'}$ yields
 the Hausdorff dimension of the set of points
 of type $(\alpha,\alpha')$. This can be generalized to the multiple scaling behavior $r^{\alpha_i},\ i=1,...,m$ of the
 potential in the $m$ sectors of an $m$-arm scaling star, with a multifractal spectrum $f_m(\{\alpha_i\})$,
 to be calculated below.\\
 
We shall use conformal tools (linked to quantum gravity),
 which allow the mathematical
description of random walks interacting with CI fractal structures,
thereby yielding a complete, albeit probabilistic,
description of the potential. The results are applied directly to
well-recognized universal fractals, like $O(N)$ loops or Potts clusters. In particular, a subtle
geometrical structure is observed in Potts clusters, where the {\it external perimeter} (EP),
which bears the electrostatic charge, differs from the full cluster's  hull. Its fractal dimension
$D_{\rm EP}$ is obtained
exactly, generalizing the case of percolation elucidated in \cite{DAA}. We obtain a duality relation:
\begin{equation}
(D_{\rm EP}-1)(D_{\rm H}-1)=\frac{1}{4}\, ,
\label{D-D}
\end{equation}
where $D_{\rm H} \geq D_{\rm EP}$ is the hull dimension. Notice that the symmetric point of (\ref{D-D}) is 
$3/2$, which is the maximum dimension of a {\it simple} (i.e., without double points) conformally invariant random curve 
in the plane. This duality, which actually gives clusters EP's as simple $O(n)$ loops at their critical point, 
was first obtained in \cite{duplantier11}. It predicts a corresponding duality in the $SLE$ process (see below).

The {\it quantum gravity} techniques used
here are not yet widely
known in statistical mechanics, since they originally belonged to string or random matrix theory.
These techniques, moreover, are not yet within the realm of
rigorous mathematics. Contact will be made with rigorous
results recently obtained in probability theory for
Brownian motion and conformally invariant scaling curves \cite{lawler2,lawler3,schramm1,lawler4,lawler5},
or percolation\cite{smirnov1,lawler6,smirnov2}, which, by completely different techniques
(using in particular the so-called
``stochastic L\"owner evolution'' ($SLE$) \cite{schramm1}), parallel those of statistical mechanics and quantum gravity.
In particular, our duality equation (\ref{D-D}) brings in the $\kappa\kappa'=16$ duality, where the 
$SLE_{\kappa'}$ process is the simple frontier of the $SLE_{\kappa}$, for $\kappa\geq 4$. This of course hints
at deep connections between  probability theory
and conformal field theory. In particular, the correspondence extensively used here,
which exists between scaling laws in the plane, and on a random Riemann surface appears as
fundamental.\\

\textbf{\large{\textbf{II Intersections of Random Walks}}}\\

Let us first define intersection exponents for random walks or Brownian motions, which, while simpler
than the multifractal exponents considered above, in fact generate the latter. Consider a number $L$ of independent random
walks (or Brownian paths) $B^{(l)},l=1,..,L$ in $\bf Z^{d}$ (or $\bf R^{d}),$ 
starting at
fixed neighboring points, and the probability 
\begin{equation}
P_{L}\left( t\right) =P\left\{ 
\cup^{L}_{l, l'=1} (B^{(l)}\lbrack 0,t\rbrack \cap B^{(l')}\lbrack 0,t\rbrack) 
=\emptyset
\right\},
\label{pl}
\end{equation} 
that the intersection of their paths up to time $t$ is 
empty\cite{lawler1,duplantier1}. At
large times and for $d<4,$ one expects this probability to decay as 
\begin{equation}
P_{L}\left( t\right) \approx t^{-\zeta_{L}},
\label{zeta}
\end{equation} 
where $\zeta _{L}\left( d\right) $
is a {\it universal} exponent depending 
only on $L$ and $d$. Above the upper critical  dimension $d=4$, RWs  almost 
surely do not intersect. The existence of exponents $\zeta
_{L}$ in $d=2, 3$ and their universality 
have been proven\cite{burdzy}, and they can be calculated near $%
d=4$ by renormalization theory \cite{duplantier1}. A generalization was introduced 
\cite{duplantier2} for $L$ walks constrained to 
stay in
a half-plane, and starting at neighboring points on the boundary, with a 
non-intersection
probability $\tilde{{P}_{L}}\left( t\right) $ of their paths governed by a
``surface'' critical exponent $\tilde{\zeta}_{L}$ such that
\begin{equation} 
\tilde{P}
_{L}\left( t\right) \approx t^{-\tilde{\zeta}_{L}}.
\label{zetat}
\end{equation}  

It was conjectured from conformal invariance arguments and numerical
simulations that in 2D \cite{duplantier2}
\begin{equation}
\zeta _{L}=h_{0, L}^{\left( c=0\right) }=\frac{1}{24}\left( 4L^{2}-1\right), 
\label{Zeta}
\end{equation}
and for the half-plane
\begin{equation}
2\tilde{\zeta}_{L}=h_{1, 2L+2}^{\left( c=0\right) }=\frac{1}{3}L\left(
1+2L\right), 
\label{C2}
\end{equation}

where $h_{p,q}^{(c)}$ denotes the Ka\v {c} conformal weight 
\begin{equation}
h_{p,q}^{(c)}=\frac{\left[ (m+1)p-mq\right] ^{2}-1}{4m\left( m+1\right) },
\label{K}
\end{equation}
of a minimal conformal field theory of central charge $c=1-6/m\left(
m+1\right) ,$ $m\in \bf N^{*}$ \cite{friedan}. For Brownian motions $%
c=0,$ and $m=2.$ For $L=1,$ the intriguing $\zeta _{1}=1/8$ is actually 
the disconnection exponent governing the probability that the origin of a
single walk remains accessible from infinity without crossing the walk.

To derive the conjectured intersection exponents above, the idea \cite{duplantier7} is to map the original
random walk problem in the plane onto a random lattice with planar geometry, or, 
in other words,
in presence of two-dimensional {\it quantum gravity} \cite{polyakov}. The key point 
is that the random walk intersection exponents on the random lattice are related 
to those in the plane. Furthermore, the RW intersection problem can be solved in 
quantum gravity. Thus, the exponents $\zeta_{L}$ (Eq. (\ref{Zeta})) and $\tilde{\zeta}_{L}$ (Eq. (\ref{C2}) in the 
standard Euclidean plane are derived from this mapping to a random lattice or Riemann surface. 

Random 
surfaces, in relation to string theory \cite{see2}, have been the subject and 
source of important developments in statistical
mechanics in two-dimensions. In particular, the discretization
of string models led to the consideration of abstract random lattices $G$, the 
connectivity fluctuations of which represent those of
the metric, i.e.~pure 2D quantum gravity \cite{boulatov}.
One can then put any 2D statistical model (like Ising model 
\cite{kazakov}, self-avoiding walks \cite{DK}) on the random planar graph 
$G$, 
thereby obtaining a new critical behavior, corresponding to the confluence of 
the
criticality of the random surface $G$ with the
critical point of the original model. The critical system ``dressed by gravity'' 
has a
larger conformal symmetry which allowed Knizhnik, Polyakov, and Zamolodchikov
(KPZ) \cite{polyakov,david2} to establish the existence of a {\it relation} between the
conformal dimensions $\bigtriangleup ^{\left( 0\right) }$ of scaling
operators in the plane and those in presence of gravity, $\bigtriangleup$ :
\begin{equation}
\bigtriangleup ^{\left( 0\right) }=\bigtriangleup \left(
\bigtriangleup -\gamma \right) /(1-\gamma),
\end{equation}
where $\gamma$ is a parameter related 
to the central charge of the statistical
model in the plane: 
\begin{equation}
c=1-6\gamma^{2}/\left(1-\gamma\right); 
\end{equation}
for a minimal 
model
of the series~(\ref{K}), with $\gamma=-1/m$, and $\bigtriangleup _{p,q}^{\left(
0\right) }\equiv h_{p,q}^{\left( c\right) }.$

Let us now consider as a statistical model {\it random walks} on a {\it random
graph}. We know \cite{duplantier2} that their central charge $c=0$, whence 
$m=2$, $\gamma=-1/2.$ Thus the KPZ relation becomes
\begin{equation} 
\bigtriangleup ^{\left( 0\right) }={U}\left( \Delta 
\right)\equiv\frac{1}{3}\bigtriangleup \left( 
1+2\bigtriangleup
\right), 
\label{KPZ}
\end{equation}
which has exactly the same analytical form as the conjecture (\ref{C2})! Thus, from the KPZ equation
one infers that the planar Brownian intersection exponents
Eqs. (\ref{Zeta},\ref{C2}) are equivalent to Brownian intersection exponents in quantum gravity:
\begin{equation}
{\triangle }_{L}=\frac{1}{2}(L-\frac{1}{2}),
\label{delta}
\end{equation}
\begin{equation}
\tilde{\triangle }_{L}=L.
\label{deltatilde}
\end{equation}
Let us now sketch the derivation of the latter quantum gravity exponents\cite{duplantier7}.

Consider the set of planar random graphs $G$, built up with,
e.g., trivalent vertices tied together in a {\it random way}. The topology is fixed here to be that of a sphere $%
\left( {\cal S}\right) $ or a disc $\left( \mathcal {D}\right) $. 
The partition function is
defined as 
\begin{equation}
Z_{\chi}(\beta )=\sum _G{1\over S(G)}e^{-\beta
\left| G\right| }, 
\label{Zchi1}
\end{equation}
where $\chi $ denotes the Euler characteristic $\chi=2\left( {\cal S}\right)
,1\left( {\cal D}\right) ;\left| G\right|$ is the number of vertices of $G$, 
$S\left(
G\right) $ its symmetry factor. The partition sum converges for all values
of the parameter $\beta $ larger than some critical $\beta_c$. At $\beta
\rightarrow \beta_c^{+},$ a singularity appears due to the presence of
infinite graphs in (\ref{Zchi1}) 
\begin{equation}
Z_{\chi}\left( \beta \right) \sim \left( \beta -\beta_c\right) ^{2-\gamma
_{\rm str}(\chi)}, 
\label{Zchi2}
\end{equation}
where $\gamma _{\rm str}(\chi)$ is the string susceptibility exponent. For pure
gravity as described in (\ref{Zchi1}), the embedding dimension $d=0$
coincides with the central charge $c=0,$ and $\gamma _{\rm 
str}(\chi)=2-\frac{5}{4}%
\chi $\cite{kostov}.
 
 Now, put a set of $L$ random walks ${\cal B}=\{{B}_{ij}^{\left( l\right) 
},l=1,...,L\}$ 
on the {\it random graph} $G$ with the
special constraint that they start at the same vertex $i\in G,$ end at the same
vertex $j \in G$, and have no intersections in between. We introduce the $L-$walk 
partition function on the random lattice \cite{duplantier7}: 
\begin{equation}
Z_L(\beta ,z)=\sum _{{\rm planar}\ G}{1\over S(G)}
e^{-\beta \left| G\right| }\sum _{i,j\in G}
\sum_{\scriptstyle B^{(l)}_{ij}\atop\scriptstyle l=1,...,L}
z^{\left| {\cal B}\right| },
\label{Zl}
\end{equation}
where a fugacity $z$ is associated with the total
number $\left| {\cal B}\right| =\left| \cup^{L}_{l=1} B^{(l)}\right| $
of vertices visited by the walks. 

We generalize this to the {\it boundary} case where $G$ now has the topology of 
a disc and where the random walks connect two sites $i$ and $j$ now on the
boundary $\partial G:$%
\begin{equation}
\tilde Z_{L}(\beta , {\beta}^{\prime}, z)=\sum _{ {\rm disc}\ G}
e^{-\beta \left| G\right| }e^{-{\beta}^{\prime}{\left| \partial G\right|}} 
\sum _{{i,j} \in G}\sum_{\scriptstyle
B^{(l)}_{ij}\atop\scriptstyle l=1,...,L}
z^{\left|{\cal B}\right| },
\label{Ztilde}
\end{equation}
where $e^{-{\beta}^{\prime}}$ is the fugacity associated with the boundary's 
length. 

The double grand canonical partition function (\ref{Zl}) associated with 
non-intersecting RW's on a random lattice can be calculated exactly 
\cite{duplantier7}.  
The critical behavior of $Z_{L}\left( \beta ,z\right)$ is then 
obtained by taking the double
scaling limit $\beta \rightarrow \beta_c$ (infinite random surface) and $%
z\rightarrow z_c$ (infinite RW's). 
The analysis of this singular behavior in terms of conformal dimensions is 
performed by using {\it finite size scaling} (FSS) \cite{DK}, where one 
must have  $\left| {\cal B}\right| \sim \left| G\right| 
^\frac{1}{2}$. One obtains \cite{duplantier7}:
 
\begin{equation}
Z_{L}\left( \beta ,z\right) \sim \left( \beta -\beta_c\right) ^{L} \sim {\left| 
G\right|} 
^{-L}.
\label{Zll}
\end{equation}
$Z_{L}$ (\ref{Zl}) represents a random surface with two {\it punctures} where 
two conformal operators of dimension $\bigtriangleup _{L}$ are
located (here
two vertices of $L$ non-intersecting RW's), and in a graphical way scales as
\begin{equation}
Z_{L}\sim Z\lbrack 
\hbox to 8.5mm{\hskip -3mm 
$\vcenter{
\epsfig{file=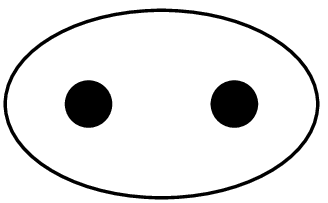, height=4.5mm}}$ 
              \hskip -80mm}
 \rbrack \ \times \left| G\right|
^{-2\triangle _{L}}\
\label{Zls}
\end{equation}
where the partition function of the two-puncture surface is the second derivative of 
$Z_{\chi=2 }(\beta)$ (\ref{Zchi2}). The latter two equations yield
\begin{equation}
2\triangle _{L}-\gamma _{\rm str}(\chi =2)=L,
\label{deltal}
\end{equation}
where $\gamma _{\rm str}(\chi =2)=-1/2$. We thus get the announced result
\begin{equation}
\triangle _{L}=\frac{1}{2}(L-\frac{1}{2}).
\label{deltaL}
\end{equation}

For the boundary partition function $\tilde Z_{L}$
(\ref{Ztilde}) a similar analysis can be performed near the triple critical 
point where the boundary length also diverges. The boundary 
partition
function $
\tilde Z_{L}$ corresponds to two boundary operators of dimensions 
${\tilde \triangle}_{L},$ integrated over $\partial G,$ on a
random surface with the topology of a disc, or in graphical terms: 
\begin{equation}
\tilde Z_{L}\sim Z(
\hbox to 9.5mm{\hskip -3mm 
$\vcenter{\epsfig{file=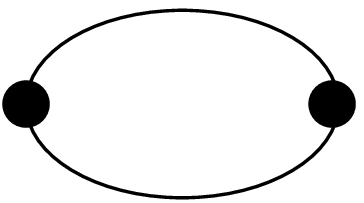, height=4.5mm}}$ 
              \hskip -80mm}
) \times \left|\partial
G\right| ^{-2\tilde{\triangle }_{L}}.  
\label{Zlts}
\end{equation}
From the exact calculation of the {\it boundary} partition function 
(\ref{Ztilde}), one gets the further equivalence to the {\it bulk} one:
\begin{equation}
\tilde Z_{L}/Z(
\hbox to 9.5mm{\hskip -3mm 
$\vcenter{\epsfig{file=fig3.eps, height=4.5mm}}$ 
              \hskip -80mm}
) 
\sim Z_{L}
,
\label{ratio}
\end{equation}
where the equivalences hold true in terms of scaling
behavior. Comparing eqs.~(\ref{Zlts})~(\ref{ratio}), and~(\ref{Zll}), 
and using the FSS $\left| \partial
G\right|\sim \left| G\right| ^{1/2}$ gives \begin{equation}
{\tilde \triangle}_{L}=L. 
\label{deltat}
\end{equation}
Applying the quadratic KPZ relation (\ref{KPZ}) to $\triangle
_{L}$
and ${\tilde \triangle}_{L}$ above yields at once the values
in the plane ${\bf R}^{2},\triangle _{L}^{(0)}\equiv \zeta
_{L}$ (Eq. (\ref{Zeta})), and ${\tilde \triangle}_{L}^{(0)}\equiv 2{\tilde 
\zeta}_{L}$ (Eq. (\ref{C2})).

Consider
now the exponents $\zeta 
(n_{1},..,n_{L})=\triangle ^{(0)}
\left\{ n_{l}\right\},$ as well as $2{\tilde \zeta}(n_{1},..,n_{L}) ={\tilde %
\triangle}^{(0)}\left\{ n_{l}\right\} ,$ describing $L$
mutually-avoiding bunches $l=1,..,L$, each made of $n_{l}$ {\it independent} walks, i.e., mutually ``transparent''\cite{werner}, with possible mutual intersections 
in a bunch. In presence of gravity each bunch will contribute 
its own {\it normalized boundary partition function} as a factor, and yield a 
natural generalization of~(\ref{ratio})

\begin{equation}
Z\left\{ n_{l}\right\} \sim {{\tilde Z}\left\{ n_{l}\right\} \over Z(
\hbox to 9.5mm{\hskip -3mm 
$\vcenter{\epsfig{file=fig3.eps, height=4.5mm}}$ 
              \hskip -80mm}
)}
\sim \prod _{l=1}^{L}
{{{\tilde Z}\left( n_{l}\right)\over Z(
\hbox to 9.5mm{\hskip -3mm 
$\vcenter{\epsfig{file=fig3.eps, height=4.5mm}}$ 
              \hskip -80mm}
)}
}, 
\label{Zn}
\end{equation}
to be identified with $\left|\partial G\right| ^{-2{\tilde \triangle}%
\left\{ n_{l}\right\}}$. The {\it factorization} property (\ref{Zn}) immediately
implies the {\it additivity of boundary conformal dimensions in presence of 
gravity} 
\begin{equation}
{\tilde \triangle}\left\{n_{1},..,n_{L}\right\} =\sum ^{L}_{l=1}
{\tilde \triangle}(n_{l}),
\label{deltan}
\end{equation}
where ${\tilde \triangle}(n)$ is now the boundary dimension of
a {\it single} bunch of $n$ transparent walks on the random surface.
We know ${\tilde \triangle}(n)$ exactly since it
corresponds in the standard plane to a trivial surface conformal dimension 
${\tilde \triangle}^{(0)}(n)=n.$ It
thus suffices to {\it invert} (\ref{KPZ}) to get 
\begin{equation}
{\tilde \triangle}(n)={U}^{-1}(n)=\frac{1}{4}(\sqrt{24n+1}-1). 
\label{deltatn}
\end{equation}
One notes the identification (\ref{Zn}), on a random surface,  
of the bulk partition function with the ratio of boundary ones. 
In the plane, using once again the KPZ relation (\ref{KPZ}) for ${\tilde 
\Delta}\left\{ n_{l}\right\} $ gives the general 
results \cite{duplantier7}
\begin{eqnarray}
\zeta (n_{1},..,n_{L})  &=&V(x)\equiv\frac{1}{24} (4x^{2}-1),
\label{Zetal}
\end{eqnarray}
\begin{eqnarray}
2{\tilde \zeta}(n_{1},..,n_{L}) =U(x)=\frac{1}{3} x(1+2x),
\label{Zetall}  
\end{eqnarray}
\begin{eqnarray}
x =\sum_{l=1}^{L}{U}^{-1}(n_{l})=\sum_{l=1}^{L}\frac{1}{4}(\sqrt{24n_{l}+1}-1). \text{ } 
\label{ZetaL}
\end{eqnarray}
Lawler and Werner \cite{lawler2} proved by probabilistic means, using the geometrical conformal invariance of Brownian motions,
the {\it existence} of two functions $U$ and $V$ satisfying the structure (\ref{Zetal}-\ref{ZetaL}). The quantum gravity 
approach here understands this
structure in terms of linear equation (\ref{deltan}), 
and yields the explicit functions $U(x)$ and $V(x)\equiv 
U(\frac{1}{2}(x-\frac{1}{2}))$ of (\ref{Zetal})(\ref{Zetall}). The same expression of these functions
has been finally derived in probability theory\cite{lawler4}.

Let us remark that the above equations yield for $\zeta (2,1^{(
L)})$ describing a two-sided walk and $L$ one-sided walks, all
mutually non-intersecting,
\begin{eqnarray}
\zeta (2,1^{(L)})= \zeta_{L+\frac{3}{2}}=V(L+\frac{3}{2})
=\frac{1}{6}(L+1)(L+2).
\end{eqnarray}
\label{Mand}
For $L=1,\zeta (2,1)=\zeta _{5/2}=1$ gives correctly the escape probability of a 
RW from another RW. For $L=0,\zeta 
(2,1^{(0)})=\zeta _{3/2}=1/3$ is related to the Hausdorff dimension of the
frontier by $D=2-2\zeta$\cite{lawler}. Thus we obtain \cite{duplantier7}
\begin{equation}
D=2-2{\zeta}_{\frac{3}{2}}=\frac{4}{3}, 
\label{mand}
\end{equation}
i.e., 
{\it Mandelbrot's
conjecture}. 
This conjecture has been finally proven in probability theory \cite{lawler5}, using 
the analytic properties of the exponents derived from the so-called stochastic L\"owner 
equation \cite{schramm1}.  
The quantum geometric structure explicited here allows
several generalizations, which we now describe \cite{duplantier8}.\\

\textbf{\large{\textbf{III Random Walks and Self-avoiding Walks}}}\\

We now generalize the scaling structure obtained in the preceding section to arbitrary sets of random or
sel-avoiding walks \cite{duplantier8} (see also \cite{lawler2,lawler3}).
Consider a general star copolymer ${\cal S}$ in the plane ${\bf R}^{2}$ (or
in ${\bf Z}^{2}$), made of an arbitrary mixture of Brownian paths or RW's $%
\left( \text{set}{\rm \;}{\cal B}\right) ,$ and polymers or SAW's $\left( 
\text{set}{\rm \;}{\cal P}\right) ,$ all starting at neighboring points. Any
pair $\left( A,B\right) $ of such paths, $A,B\in {\cal B}$ or ${\cal P},$
can be constrained in a specific way: either they avoid each other $\left(
A\cap B=\emptyset ,\text{ noted }A\wedge B\right) ,$ or they are independent, i.e., ``transparent''
and can cross each other (noted $A\vee B)$\cite{duplantier8,ferber}. This notation allows any {\it 
nested}
interaction structure \cite{duplantier8}; one can decide for instance that the 
branches $%
\left\{ P_{\ell }\in {\cal P}\right\} _{\ell =1,...,L}$ of an $L$-star
polymer, all mutually-avoiding, further avoid a bunch of Brownian paths $%
\left\{ B_{k}\in {\cal B}\right\} _{k=1,...,n},$ all transparent to each
other: 
\begin{equation}
{\cal S}=\left( \bigwedge\nolimits_{\ell =1}^{L}P_{\ell }\right) \wedge
\left( \bigvee\nolimits_{k=1}^{n}B_{k}\right) .  \label{vw}
\end{equation}
In 2D the order of the branches of the star copolymer {\it %
does} matter and is intrinsic to our $\left( \wedge ,\vee \right) $ notation.

To each {\it specific} star copolymer center ${\cal S}$ is attached a
conformal scaling operator with a scaling dimension $x\left( {\cal S}\right)
.$ To obtain proper scaling we consider the partition functions of Brownian paths 
and polymers having the same mean size $R$. When the star is constrained to stay 
in a {\it half-plane} with its core
placed near the {\it boundary}, its partition function will scale with new 
boundary scaling dimension $\tilde{x}\left( {\cal S}\right)$ \cite
{duplantier2,duplantier4,DS2}.

Any scaling dimension $x$ in the bulk is twice the {\it conformal dimension}
(c.d.) $\Delta ^{(0)}$ of the corresponding operator, while near a boundary
(b.c.d.) they are identical: 
\begin{equation}
x=2\Delta ^{\left( 0\right) },\quad \tilde{x}=\tilde{\Delta}^{\left(
0\right) }.  \label{xdelta}
\end{equation}

As above, the idea is to use the representation where the RW's or SAW's are on a 
2D
random lattice, or a random Riemann surface, i.e., in the presence of 2D {\it 
quantum gravity} \cite
{polyakov}. The general relation (\ref{KPZ}) depends only on the central charge, 
and is valid for polymers, for which $c=0$. Let us summarize the results 
\cite{duplantier8}, expressed here in terms of the scaling dimensions in the 
standard plane. For a critical system with central charge $c=0$, the two universal 
functions:
\begin{eqnarray}
U\left( x\right) &=&\frac{1}{3}x\left( 1+2x\right) , \hskip2mm V\left( x\right) 
=\frac{1}{24}\left( 4x^{2}-1\right) ,  \label{U}
\end{eqnarray}
with $V\left( x\right) \equiv U\left( 
{\textstyle{1 \over 2}}%
\left( x-%
{\textstyle{1 \over 2}}%
\right) \right),$
generate all the scaling exponents. The scaling exponents $x\left( A\wedge
B\right) $, and $\tilde{x}\left( A\wedge B\right) ,$ of two {\it mutually 
avoiding} stars $A,B,$ with proper scaling 
exponents $x\left( A\right) ,x\left( B\right) ,$ or boundary exponents 
$\tilde{x}\left( A\right) ,\tilde{x}\left(
B\right) ,$  
obey the {\it star
algebra} \cite{duplantier7,duplantier8} 
\begin{eqnarray}
x\left( A\wedge B\right) &=&2V\left[ U^{-1}\left( \tilde{x}\left( A\right)
\right) +U^{-1}\left( \tilde{x}\left( B\right) \right) \right]  \nonumber \\
\tilde{x}\left( A\wedge B\right) &=&U\left[ U^{-1}\left( \tilde{x}\left(
A\right) \right) +U^{-1}\left( \tilde{x}\left( B\right) \right) \right] ,
\label{x}
\end{eqnarray}
where $U^{-1}\left( x\right) $ is the inverse function of $U$
\begin{equation}
U^{-1}\left( x\right) =\frac{1}{4}\left( \sqrt{24x+1}-1\right) .  \label{u1}
\end{equation}
On a random surface, $U^{-1}\left( \tilde{x} \right)$ is the boundary
dimension corresponding to the value $\tilde{x}$ in ${\bf R} \times
{\bf R}^{+}$, and the sum of $U^{-1}$ functions in Eq. (\ref{x})
represents linearly the juxtaposition $A \wedge B$ of two sets of
random paths near their random frontier, i.e., the product of two
``boundary operators'' on the random surface. The latter sum is mapped
by the functions $U$, $V$, into the scaling dimensions in ${\bf
R}^2$ \cite{duplantier8}. 

The rules (\ref{x}), which mix bulk and boundary exponents, come from simple 
factorization properties on a random Riemann surface, i.e., in quantum gravity 
\cite{duplantier7,duplantier8}, (and are also 
recurrence relations in ${\bf R}^{2}$ between conformal Riemann maps of the 
successive mutually-avoiding paths onto the line ${\bf R}$\cite{lawler2}). 

If, on the contrary, $A$ and $B$ are {\it independent} and can overlap, then
by trivial factorization of probabilities their dimensions are 
additive\cite{duplantier8}
\begin{eqnarray} 
x\left( A\vee B\right)
=x\left( A\right) +x\left( B\right) ,\nonumber \\ 
\tilde{x}\left( A\vee B\right) =%
\tilde{x}\left( A\right) +\tilde{x}\left( B\right).
\label{add}
\end{eqnarray}

It is clear at this stage that the set of equations above is {\it complete.}
It allows for the calculation of any conformal dimensions associated with a star 
structure ${\cal S}$ of the most general type, as in (%
\ref{vw}), involving $\left( \wedge ,\vee \right) $ operations separated by
nested{\it \ }parentheses \cite{duplantier8}.

{\it Brownian-polymer exponents: }The single extremity scaling dimensions
are for a RW or a SAW near a Dirichlet boundary in ${\bf R}^{2}$ \cite
{DS2,cardy} 
\begin{equation}
\tilde{x}_{B}\left( 1\right)=\tilde{\Delta}_{B}^{\left( 0\right) }\left( 1\right) 
=1,\;\tilde{x}_{P}\left( 1\right)=\tilde{\Delta}%
_{P}^{\left( 0\right) }\left( 1\right) =%
{\textstyle {5 \over 8}}%
,  \label{num}
\end{equation}
or on $G,$ $\tilde{\Delta}_{B}\left( 1\right)
=U^{-1}\left( 1\right) =1,\;\tilde{\Delta}_{P}\left( 1\right) =U^{-1}\left( 
{\textstyle {5 \over 8}}%
\right) =%
{\textstyle {3 \over 4}}%
.$ Because of the star algebra described above these are the only numerical
seeds, i.e., generators, we need.

Stars can include bunches of $n$ copies of transparent RW's or $m$
transparent SAW's. Their b.c.d.'s in ${\bf R}^{2}$ are respectively, by
using (\ref{add}) and (\ref{num}), $\tilde{\Delta}_{B}^{\left( 0\right)
}\left( n\right) =n$ and $\tilde{\Delta}_{P}^{\left( 0\right) }\left(
m\right) =\frac{5}{8}m,$ from which the inverse mapping to the
random surface yields $\tilde{\Delta}_{B}\left( n\right) =U^{-1}\left(
n\right) $ and $\tilde{\Delta}_{P}\left( m\right) =U^{-1}\left( 
{\textstyle {5 \over 8}}%
m\right) .$ The star made of $L$ bunches $\ell \in \left\{ 1,...,L\right\} $, each 
of them made of $n_{\ell }$ transparent RW's and of $m_{\ell }$ transparent
SAW's, and the $L$ bunches being mutually-avoiding, has planar scaling dimensions  
\begin{eqnarray*}
\tilde{\Delta}^{\left( 0\right) }\left\{ n_{\ell },m_{\ell}\right\} &=&U\left( 
\tilde{\Delta}\right) ,\;\Delta ^{\left( 0\right)
}\left\{ n_{\ell },m_{\ell }\right\} =V\left( \tilde{\Delta}%
\right) , \\
\tilde{\Delta}\left\{ n_{\ell },m_{\ell}\right\}
&=&\sum\nolimits_{\ell =1}^{L}U^{-1}\left( n_{\ell }+ 
{\textstyle {5 \over 8}}%
m_{\ell }\right) .
\end{eqnarray*}
This encompasses all previously known exponents for RW's and SAW's
\cite{duplantier2,duplantier4,DS2}. We in particular arrive at the striking {\it scaling equivalence:
a self-avoiding walk is exactly equivalent to $5/8$ of a Brownian motion}.
Similar results have been obtained in probability theory, based on the general structure of
 ``completely conformally invariant processes'', which correspond 
 exactly to  $c=0$ central charge conformal field theories \cite {lawler3,lawler4}.
 The rigorous construction of the scaling limit of SAW still eludes a rigorous approach, eventhough it is
 predicted that it corresponds to the ``stochastic L\"owner evolution'' $SLE_{\kappa}$ with
 $\kappa=8/3$, equivalent to a Coulomb gas with $g=4/\kappa=3/2$ (see below section XI).\\

\textbf{\large{\textbf{IV Conformal Multifractality and the Harmonic Measure}}}\\

The {\it harmonic measure}, i.e., the diffusion or 
electrostatic field near an equipotential fractal boundary\cite{BBE}, or, equivalently, the electric charge 
appearing on the frontier of a perfectly conducting fractal, possesses a 
self-similarity property, which
is reflected in a {\it multifractal} (Mf) behavior. Cates and Witten \cite{cates 
et witten} considered the case of the Laplacian diffusion field near a simple 
random walk, or near a self-avoiding walk. The associated exponents can be
recast as those of star copolymers made of a bunch of independent RW's
diffusing away from a generic point of the absorber. The exact 
solution to this problem in two dimensions is as follows \cite{duplantier8}. 
From a mathematical point of view, it could also be derived from
the results of refs \cite{lawler2,lawler3,lawler4,lawler5} taken altogether. 

The two-dimensional ``absorber'' ${\cal S}$ can be a
random walk, or a self-avoiding walk. The harmonic measure
${\rm H}\left( w\right) $ is the probability that another random walker (RW) launched
from 
infinity, {\it first} hits the outer ``hull's frontier'' or (accessible) frontier ${\cal F}({\cal S})$ at point $w \in {\cal F}({\cal S})$. A covering 
of $\cal F$ by balls ${\cal B}(w, a)$ of radius $a$ is centered at points $w \in 
{\cal F}/\{a\}$ forming a discrete subset ${\cal F}/\{a\}$ of $\cal F$. Let ${\rm H}({\cal F} \cap {\cal B}(w, a))$ be the 
harmonic measure of the intersection set between $\cal F$ and the ball ${\cal B}(w, a)$. The moments of $H$, averaged over all
realizations of RW's and ${\cal S}$ are defined as
\begin{equation}
{\cal Z}_{n}=\left\langle \sum\limits_{w\in {\cal F}/\{a\}}{\rm H}^{n}\left({\cal F} \cap {\cal B}(w, 
a)\right)
\right\rangle ,  \label{Z}
\end{equation}
 where $n$ can be, {\it a priori},
a real number. In the limit of large absorbers ${\cal S}$ and frontiers ${\cal F}\left( 
{\cal 
S}\right) 
$ of average size $R,$ or small covering radius $a$, i.e, $a/R \to 0$,  these moments scale as 
\begin{equation}
{\cal Z}_{n}\approx \left( a/R\right) ^{\tau \left( n\right) },  \label{Z2}
\end{equation}
where the multifractal scaling exponents 
$\tau \left(
n\right) $ encode generalized dimensions $D\left( n\right)$, $\tau \left( 
n\right) 
=\left( n-1\right) D\left( n\right) ,$
which vary in a non-linear way with $n$\cite{mandelbrot2,hentschel,frisch,halsey}.

As explained in the introduction, the harmonic multifractal spectrum $f(\alpha)$ (Eqs (\ref{ha'},\ref{ca'},\ref{alpha})) is derived as a Legendre transform of
the $\tau(n)$ function. (The existence of the harmonic multifractal spectrum $f(\alpha)$ 
for a Brownian path has been rigorously established in\cite{lawler97}.) 

By the very definition of the H-measure, $n$ independent RW's diffusing away 
from 
the absorber give a geometric representation of the $n^{th}$
moment ${\rm H}^{n},$ for $n$ {\it integer}, and convexity arguments give the whole 
continuation to real values.  When the absorber is a RW or a SAW of size $%
R,$ the site average of its moments ${\rm H}^{n}$
 is represented by a copolymer star partition function ${\cal Z}_{R}\left( 
{\cal S}_{\wedge }n\right)$, where we have
introduced the short-hand notation ${\cal S}_{\wedge }n\equiv {\cal S}\wedge
\left( \vee B\right) ^{n}$ describing the copolymer star made by the
absorber ${\cal S}$ hit by the bunch $\left( \vee B\right) ^{n}$ at the apex
only \cite{cates et
witten,duplantier8}.\ More
precisely one has
\begin{equation}
{\cal Z}_{n}\approx R^2 {\cal Z}_{R}\left( 
{\cal S}_{\wedge }n\right) 
\end{equation}
where the absorber ${\cal S}$ is either the two-RW star $B\vee B$ or the
two-SAW star $P\wedge P,$ made of two non-intersecting SAW's.\ \ Owing to 
Eq.(\ref{Z2}), we get the scaling relation 
\begin{equation}
\tau \left(
n\right) =x\left( {\cal S}_{\wedge
}n\right) -2. 
\end{equation}
  Our formalism (\ref{x}) immediately gives the scaling dimensions 
\begin{equation}
x\left( 
{\cal S}_{\wedge }n\right) =2V\left( \tilde{\Delta}\left( {\cal S}\right)
+U^{-1}\left( n\right) \right),
\end{equation} 
 where $\tilde{\Delta}\left( {\cal S}%
\right) $ is as above the quantum gravity boundary dimension of the absorber 
${\cal S}$ alone. For a RW absorber, we have $\tilde{\Delta}\left( B\vee B\right)
=U^{-1}\left( 2\right) =\frac{3}{2},$ while for a SAW $\tilde{\Delta}\left(
P\wedge P\right) =2\tilde{\Delta}_{P}(1)=2U^{-1}\left( \frac{5}{8}\right) =%
\frac{3}{2}.$ The coincidence of these two values
tells us that {\it in} 2D {\it the harmonic multifractal spectra} $f\left( \alpha
\right) $ {\it of a random walk or a self-avoiding walk are identical.} The 
calculation gives \cite{duplantier8}
\begin{equation}
\tau \left( n\right)=\frac{1}{2}\left( n-1\right) +\frac{5}{24}\left( 
\sqrt{24n+1}-5\right) , \label{tauf}\\
\end{equation}
\begin{equation}
\alpha =\frac{d\tau }{dn}\left( n\right) =\frac{1}{2}+\frac{5}{2}\frac{1}{%
\sqrt{24n+1}},  \label{alpha1}\\
\end{equation}
\begin{equation}
D\left( n\right) =\frac{1}{2}+\frac{5}{\sqrt{24n+1}+5},\quad n\in \left[ -%
{\textstyle{1 \over 24}}%
,+\infty \right) ,  \label{dna}
\end{equation}
\begin{equation}
f\left( \alpha \right) =\frac{25}{48}\left( 3-\frac{1}{2\alpha -1}\right) -%
\frac{\alpha }{24},\quad \alpha \in \left( 
{\textstyle{1 \over 2}}%
,+\infty \right) . \label{mf}
\end{equation}  
\begin{figure}
\centerline{\epsfig{file=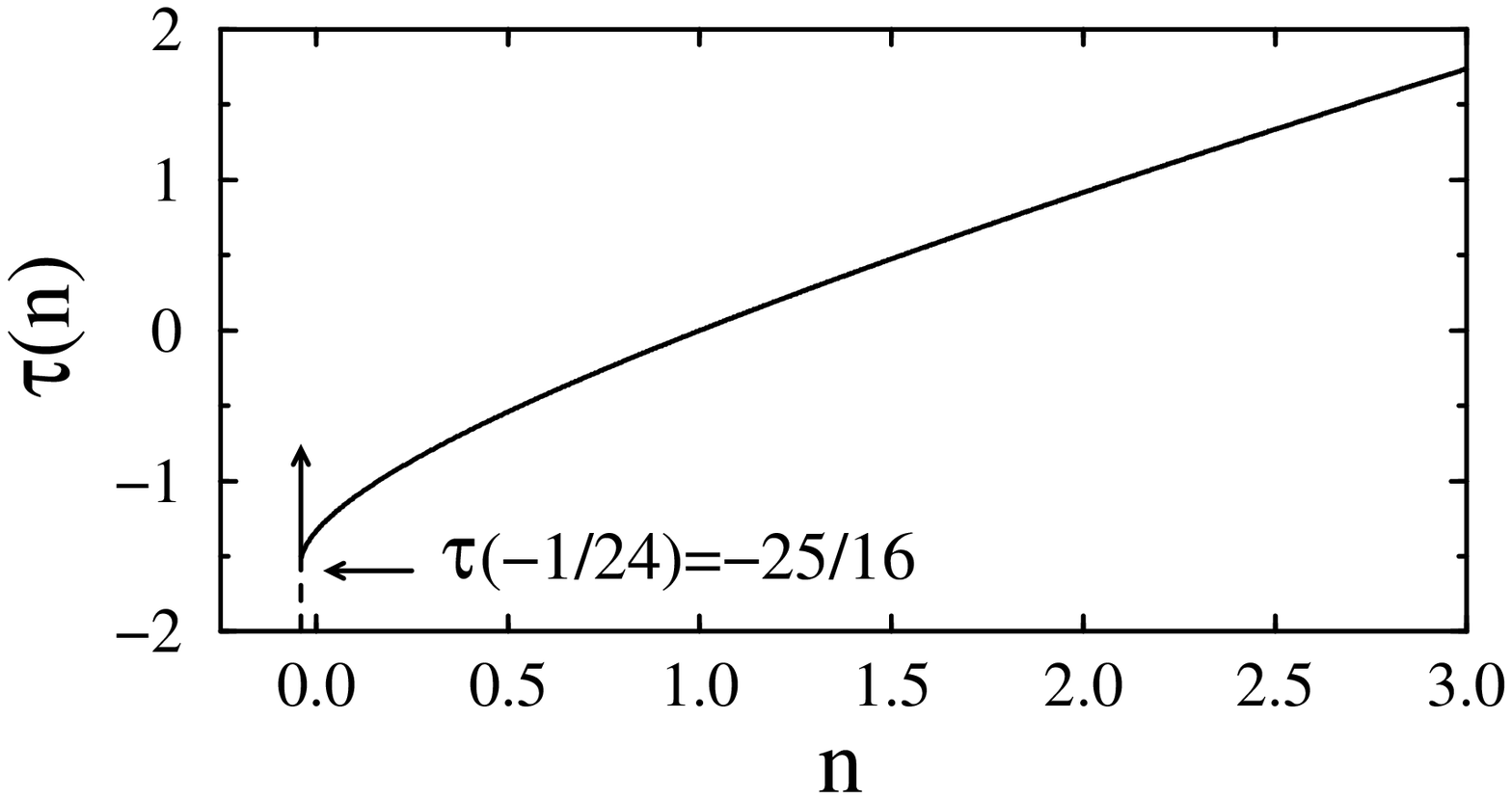,width=8.5cm}}
\end{figure}
\begin{figure}
\vskip -23pt
\centerline{\epsfig{file=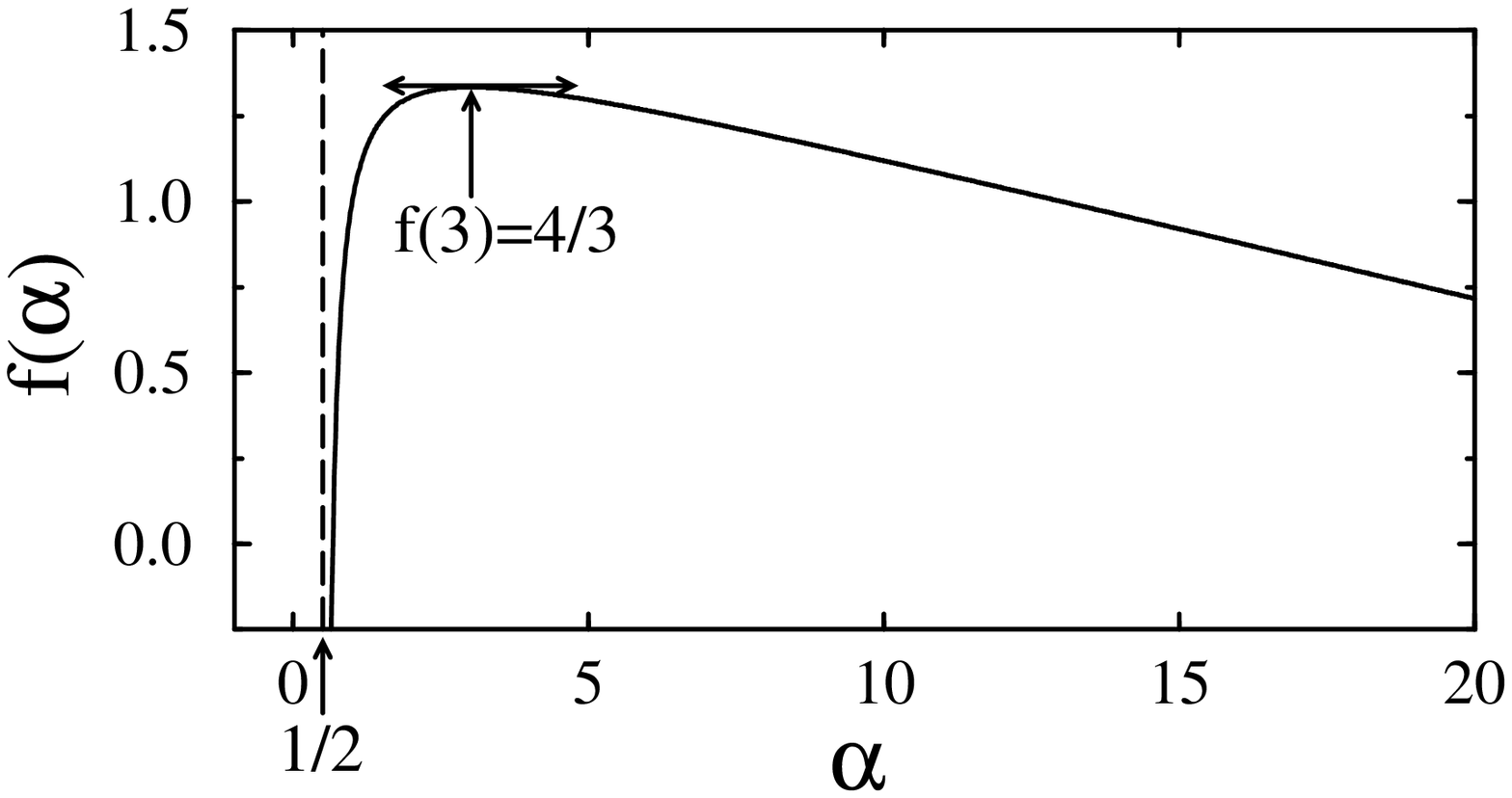,width=8.5cm}}
\smallskip
\caption{Harmonic multifractal dimensions $\tau (n)$ and spectrum
$f(\alpha)$ of a two-dimensional RW or SAW.}
\label{Figure}
\end{figure}
The corresponding universal curves are shown in Fig.1: $\tau \left( 
n\right) 
$ is half a parabola, and $f\left( \alpha \right) $ a hyperbola. $D\left(
1\right) =\tau ^{\prime }\left( 1\right) =1$ is Makarov's theorem. 
The singularity at $\alpha=\frac{1}{2}$ in the multifractal functions $f(\alpha)$ 
corresponds to points 
on the fractal boundary $\cal F$ where the latter has the local geometry of a needle. 
The mathematical version of this statement is given 
by Beurling's theorem \cite{ahlfors}, which states that at distance $\epsilon$ from the boundary, 
the harmonic measure is bounded above by
\begin{equation}
H \left(z: {\rm inf}_{w \in  {\cal F}}|z - w|  \leq \epsilon)\right) \leq C {\epsilon}^{1/2},
\label{beurling}
\end{equation}
where $C$ is a constant. This insures that the spectrum of multifractal H{\"o}lder exponents 
$\alpha$ is bounded below 
by $\frac{1}{2}$.
The right branch of the$%
f\left( \alpha \right) $ curve has a linear asymptote 
\begin{equation}
\lim_{\alpha \rightarrow +\infty} \frac{1}{\alpha}f\left(
\alpha \right) = -\frac{1}{24}. 
\end{equation}
Its linear shape is quite reminiscent of that of the multifractal function
 of the growth probability as in the case of a 2D DLA 
cluster 
\cite
{ball}. The domain of large values of $\alpha$ corresponds to the lowest part $n\rightarrow {n^{\ast}} =-\frac{1}{24}$ of the spectrum of 
dimensions, which is dominated by almost inaccessible sites, and the existence of a 
linear asymptote to the multifractal function $f$ implies a peculiar 
behavior for the number 
of those sites in a lattice setting. Indeed define ${\cal N}\left( H\right)$ as the  number of sites having a 
probability $H$ to be hit:
\begin{equation}
{\cal N}\left( H\right)={\rm Card}\left\{w \in {\cal F}: {\rm H}(w)=H \right\}.
\label{N(H)}
\end{equation}
Using the Mf formalism to change from variable $H$ to  
$%
\alpha $ (at fixed value of $a/R)$, shows that ${\cal N}\left( H\right)$ obeys, 
for
$H\rightarrow 0,$ a power law behavior
\begin{equation}
{\cal N}\left( H\right)|_{H\rightarrow 0}\approx H^{-{\tau}^{\ast}}
\label{nh}
\end{equation}  
with an exponent 
\begin{equation}
\tau ^{\ast }=1+%
\mathrel{\mathop{\lim }\limits_{\alpha \rightarrow +\infty }}%
\frac{1}{\alpha }f\left( \alpha \right)=1+n^{\ast}.  
\label{to}
\end{equation} 
Thus we predict
\begin{equation}
   \tau 
^{\ast}=\frac{23}{24}. 
\label{toc}
\end{equation}

One remarks that $-\tau \left(
0\right) =\sup_{\alpha }f\left( \alpha \right) =f\left( 3 \right) =\frac{4}{3}$ 
is the
Hausdorff dimension of the {\it Brownian frontier} or of a SAW. Thus Mandelbrot's
classical conjecture identifying the latter two is derived and generalized to the {\it whole} $f\left( \alpha \right) $ harmonic spectrum.\\

\noindent {\bf - An Invariance Property of $f(\alpha)$}\\

The expression of $f(\alpha)$ simplifies if one considers the combination:
\begin{eqnarray}
f\left( \alpha \right)-\alpha&=& \frac{25}{24}
\left[1-\frac{1}{2}\left(2\alpha -1 + \frac{1}{2\alpha -1}\right)\right].
\label{f-a}
\end{eqnarray}
Thus the multifractal function possesses the invariance symmetry \cite{BDH}
\begin{eqnarray}
f\left( \alpha \right)-\alpha=f\left( {\alpha}^\prime 
\right)-{\alpha}^{\prime},
\label{inv}
\end{eqnarray}
for $\alpha$ and ${\alpha}^{\prime}$ satisfying the duality relation:
\begin{eqnarray}
(2\alpha-1)(2{\alpha}^{\prime}-1)=1,
\end{eqnarray}
or, equivalently 
\begin{eqnarray}
{\alpha}^{-1}+{{\alpha}^{\prime}}^{-1}=2.
\label{aa'}
\end{eqnarray}
When associating a wedge angle $\theta=\pi / \alpha$ to each local singularity 
exponent $\alpha$, 
one recovers the complementary rule for angles in the plane \cite{BDH}
\begin{eqnarray}
\theta+{\theta}^{\prime}=\frac{\pi}{\alpha}+\frac{\pi}{{\alpha}^{\prime}}=2\pi.
\label{tetateta'}
\end{eqnarray}
It is interesting to note that, owing to the explicit forms (\ref{alpha1}) of 
$\alpha$
and (\ref{dna}) of $D(n)$, the condition (\ref{aa'}) reads also after a 
little algebra
\begin{equation}
D(n)+D(n')=2.
\label{DD'}
\end{equation}
This basic symmetry (\ref{inv}) reflects that of the cluster boundary itself 
under the {\it exchange of 
interior and exterior domains} (ref.\cite{BDH}).\\

\noindent {\bf - Higher multifractality of Brownian motion and self-avoiding walk}\\

It is interesting to note that one can also define {\it higher multifractal} spectra as those
depending on several $\alpha$ variables \cite{duplantier10}. A first exemple is given by the double moments
of the harmonic measure on {\it both} sides of a random fractal, taken here as a Brownian motion or
a self-avoiding walk. (The general case will be further described below in section VI.)
The fractal boundary has to be reached from both sides, so it must be a {\it simple} curve without double
points, which is naturally the case
of a SAW. For a Brownian motion, one can consider the subset of the {\it pinching} or {\it cut points}, of
Hausdorff dimension $D=2-2\zeta_{2}=3/4$, where the
path splits into two non-intersecting parts. Locally the Brownian path then is accessible from
both directions.\\
Let us define:
\begin{equation}
{\cal Z}_{n,n'}=\left\langle \sum\limits_{w\in {\cal F}/\{a\}}\left[{\rm H}_{+}(w)\right]^{n}\left[{\rm H}_{-}(w)\right]^{n'}\right\rangle,  
\label{ZZ'}
\end{equation}
where ${\rm H}_{+}(w) \equiv {\rm H}_{+}\left({\cal F} \cap {\cal B}(w, 
a)\right)$ and ${\rm H}_{-}(w)\equiv {\rm H}_{-}\left({\cal F} \cap {\cal B}(w,a)\right)$ are respectively the harmonic measures on ``left''or ``right'' sides of the random fractal. These moments have a 
multifractal scaling behavior
 \begin{equation}
{\cal Z}_{n}\approx \left( a/R\right) ^{\tau_2(n,n') },  \label{ZZ2'}
\end{equation}
where the exponents $\tau_2(n,n')$ now depend on two moment orders $n$ and $n'$. The generalization of the Legendre transform 
Eq. (\ref{alpha}) reads
\begin{eqnarray}
\alpha &=&\frac{\partial\tau_2 }{\partial n}\left( n,n'\right) , \quad \alpha' =
\frac{\partial\tau_2 }{\partial n'}\left( n,n'\right), \nonumber \\ 
f_2\left( \alpha, \alpha' \right) &=&\alpha n+\alpha' n'-\tau_2(n,n'), \nonumber \\
n&=&\frac{\partial f_2}{\partial \alpha}\left( \alpha, \alpha'
\right) , \quad n'=\frac{\partial f_2}{\partial \alpha' }\left( \alpha, \alpha'
\right).  
\label{alpha''}
\end{eqnarray} 
We find the $\tau$ exponents from the star algebra (\ref{x}):
\begin{equation}
\tau_2(n,n')=2V\left(
a'+U^{-1}\left( n\right)+U^{-1}\left( n'\right) \right)-2,
\label{taunn'}
\end{equation}
where $a'$ corresponds to the quantum gravity scaling dimension of the fractal set, i.e., the simple curve or
 the pinching point set, where the harmonic measure is evaluated on both sides.
For a Brownian motion, pinched into two parts separated by the two sets of auxiliary Brownian motions, 
representing the moments of the harmonic mesures, we so have:
\begin{equation}
a'= \tilde{\Delta}\left(
B\wedge B\right)=2\times \tilde\Delta_{B} (1)=2U^{-1}\left( 1\right)=2.
\end{equation}
 For a self-avoiding walk made of two mutually-avoiding one-sided arms, we have 
\begin{equation}
a'= \tilde{\Delta}\left(
P\wedge P\right)=2 \times \tilde{\Delta}_{P}(1)=2U^{-1}\left( \frac{5}{8}\right) =\frac{3}{2}.
\end{equation}
 After performing the double Legendre transform and some calculations, we find
\begin{eqnarray}
f_2\left( \alpha, \alpha' \right)&=&2+\frac{1}{12}-\frac{1}{3}{a''}^2
{\left[1-\frac{1}{2}\left(\frac{1}{\alpha}+\frac{1}{\alpha'}\right)\right]}^{-1} \nonumber \\
& &-\frac{1}{24}\left(\alpha+\alpha'\right),
\label{faa'}
\end{eqnarray} 
\begin{equation}
{\alpha}=2 \frac{1}{\sqrt{24n+1}}\left[{a''}+\frac{1}{4}\left(\sqrt{24n+1}+\sqrt{24n'+1}\right)\right],
\end{equation}
and a similar symmetric equation for $\alpha'$. Here $a''$ has the shifted value:
\begin{eqnarray}
a''&=&a'+\gamma=a'-\frac{1}{2}\\&=&\frac{3}{2}\ ({\rm RW}),\ {\rm or}\ a''=1\ ({\rm SAW}).
\label{a''}
\end{eqnarray} 
This doubly multifractal spectrum possesses the requested properties, like
${\rm sup}_{\alpha'} f(\alpha, \alpha')=f(\alpha),$ where $f(\alpha)$ is (\ref{mf}) above.

This can be generalized to a {\it star configuration} made of $m$ random walks or $m$
self-avoiding walks, where one looks at the simultaneous behavior of the potential in each sector
between the arms of the star (see § VI below for a more precise description in the general case). The
 {\it poly-multifractal} results read for Brownian motions or self-avoiding polymers:
\begin{eqnarray}
f_m\left(\{ {\alpha}_{i=1,...,m}\}\right)&=&2+\frac{1}{12}-\frac{1}{3}{a''}^2{\left(1-\frac{1}{2}
\sum_{i=1}^{m}{\alpha}_{i}^{-1}\right)}^{-1} \nonumber \\
& &-\frac{1}{24}\sum_{i=1}^{m}{\alpha}_{i},
\label{fai}
\end{eqnarray}
with 
\begin{equation}
{\alpha}_{i}=2 \frac{1}{\sqrt{24n_{i}+1}}\left({a''}+\frac{1}{4}\sum_{j=1}^{m}\sqrt{24n_{j}+1}\right),\label{ai}
\end{equation}
and where 
\begin{equation}
a'= m,\ a''=a'-\frac{1}{4}m=\frac{3}{4}m,
\end{equation}
for $m$ {\it random walks in a star configuration}, and
\begin{equation}
a'=\frac{3}{4}m,\ a''=a'-\frac{1}{4}m=\frac{1}{2}m,
\label{a''SAW} 
\end{equation}
 for $m$ {\it self-avoiding walks in a star configuration}. The two-sided case above (\ref{a''}) is recovered for $m=2$.
 The domain of definition of the poly-multifractal function $f$ is given by
\begin{equation}
1-\frac{1}{2}\sum_{i=1}^{m}{\alpha}_{i}^{-1} \geq 0,
\end{equation}
as verified by Eq. (\ref{ai}).\\

\textbf{\large{\textbf{V Percolation Clusters}}}\\

Consider now 
a two-dimensional very large incipient cluster ${\cal C}$, at the 
percolation threshold $p_{c}$. Define 
${\rm H}\left( w\right) $ as the probability that a random walker (RW) launched from 
infinity, {\it first} hits the outer (accessible) percolation hull's frontier ${\cal 
F}({\cal C})$ at point $w \in {\cal F}({\cal C})$. The moments of $H$ are averaged 
over all
realizations of RW's and ${\cal C}$, as in Eq.(\ref{Z}) above. 
For very large clusters ${\cal C}$ and frontiers ${\cal F}\left( 
{\cal 
C}\right) 
$ of average size $R,$ one expects again these moments to scale as in Eq. 
(\ref{Z2}): ${\cal Z}_{n}\approx \left( a/R\right) ^{\tau \left( n\right) }$. 
These exponents $\tau(n)$ have been obtained recently \cite{duplantier9}, 
using an exact result on the external boundary of a percolation cluster \cite{DAA}. 

We consider site percolation 
on the 2D triangular lattice. Figure 2 depicts $n$ independent random walks, 
in a bunch, {\it %
first} hitting the external frontier of a percolation cluster at a site $w=\left(
\bullet \right) .$ This site, to belong to the {\it accessible} part of the hull, must 
remain, in the 
{\it continuous scaling limit},  the
source of at least {\it three non-intersecting crossing paths}, noted ${\cal 
S}_{3},$ 
reaching to a
(large) distance $R$ (Fig. 3) \cite{DAA}. (Notice that the definition of the {\it 
standard} hull requires only a pair of dual lines). The $n$
independent RW's, or Brownian paths ${ B}$ in the scaling limit, in a bunch 
noted $\left(
\vee {B}\right) ^{n},$ {\it avoid} the set ${\cal S}_{3 }\equiv \left( 
\wedge 
{ \cal P}\right) ^{3 }$ of three {\it non-intersecting} connected paths in 
the
percolation system, and this system is governed by a new critical exponent
 $x\left( {\cal 
S}_{3 
}\wedge n\right) $ depending on $n.$
\begin{figure}
\centerline{\epsfig{file=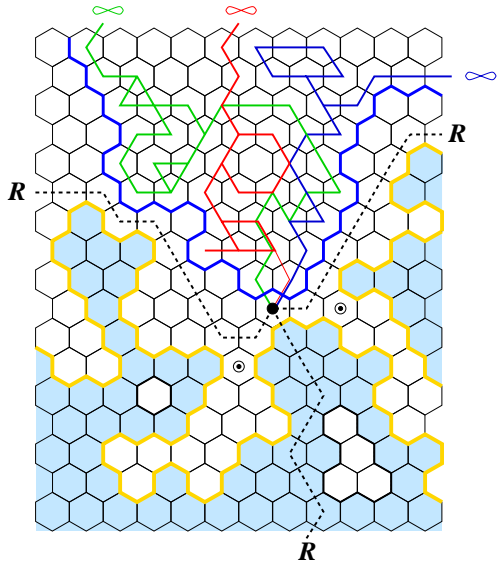,width=6.5cm}}
\smallskip
\caption{An accessible site $(\bullet )$ on the external
perimeter for site percolation on the triangular lattice. It is
defined by the existence, in the {\it scaling limit}, of three non-intersecting, 
and connected paths ${\cal S}_{3}$ (dotted
lines), one on the incipient cluster, the other two on
the dual empty sites. The entrances of fjords $\odot$ close in the scaling limit. 
Point $(\bullet )$ is first reached by three
independent RW's (red, green, blue), contributing to ${\rm H}^3 (\bullet
)$. The hull of the incipient cluster (golden line) avoids the outer
frontier of the RW's (thick blue line).}
\label{Figure2}
\end{figure} 
In terms of these definitions, the harmonic measure moments simply 
scale with an exponent \cite{duplantier8} 
\begin{equation}
\tau \left( n\right) =x\left( {\cal S}_{3}\wedge n\right) -2.  \label{tt}
\end{equation}
For percolation, two values of half-plane crossing exponents $\tilde{x}_{\ell }$
are known by
{\it elementary} means: $\tilde{x}_{2}=1,\tilde{x}_{3}=2\cite{ai1}$. We fuse the 
two objects 
${\cal 
S}_{3}$ and $\left( \vee { B}\right) ^{n}$ into a new star ${\cal 
S}_{3}\wedge \left( \vee { B}\right) ^{n}$, and use (\ref{x}) to obtain
\begin{equation}
x\left( {\cal S}_{3}\wedge n\right) =2V\left(
U^{-1}\left(\tilde{x}_{3}\right)+U^{-1}\left( n\right) \right).  \label{fina}
\end{equation}
Specifying $U^{-1}\left(\tilde{x}_{3}\right)=\frac{3}{2}$ finally gives from 
(\ref{U})(\ref{u1}) 
\[
x\left( {\cal S}_{3}\wedge n\right) =2+\frac{1}{2}\left( n-1\right) +\frac{5%
}{24}\left( \sqrt{24n+1}-5\right) .
\]
From this $\tau \left( n\right) $ (\ref{tt}) is found to be {\it identical} to 
(\ref{tauf}) for RW's and SAW's; $D\left( n\right) $ follows as: 
\begin{equation}
D\left( n\right) =\frac{1}{2}+\frac{5}{\sqrt{24n+1}+5},\quad n\in \left[ -%
{\textstyle{1 \over 24}}%
,+\infty \right) ,  \label{dn}
\end{equation}
valid for all values of moment order $n,n\geq -\frac{1}{24}.$
The Legendre transform reads again exactly as in Eq. (\ref{mf}):
\begin{equation}
f\left( \alpha \right) =\frac{25}{48}\left( 3-\frac{1}{2\alpha -1}\right) -%
\frac{\alpha }{24},\quad \alpha \in \left( 
{\textstyle{1 \over 2}}%
,+\infty \right) . \label{f}
\end{equation}

Only in the case of percolation has the harmonic measure been systematically studied numerically, 
by Meakin et al. \cite{meakin}. We give in Figure 3 the exact curve $D\left( n\right) $ (\ref{dn}) 
 \cite{duplantier9} together with the 
numerical results for $n\in \{2,...,9\} $ \cite{meakin}, showing 
fairly good agreement.

\begin{figure}
\centerline{\epsfig{file=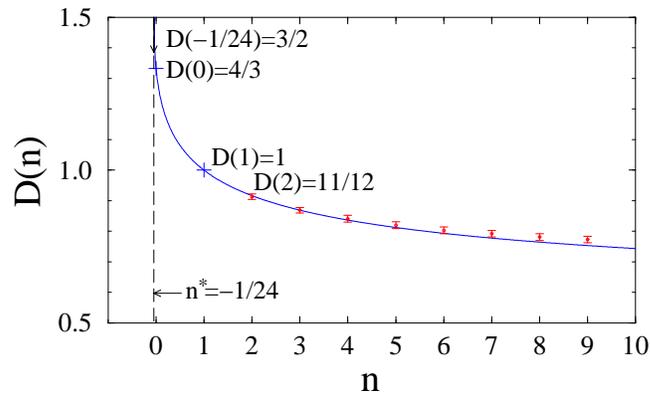,width=8.5cm}}
\medskip
\caption{Universal generalized dimensions $D(n)$ as a function of $n$, corresponding to the harmonic measure near a percolation cluster, or to self-avoiding or 
random walks, and comparison with 
the numerical data obtained by Meakin et al. (1988) for percolation. 
}
\label{Figure3} 
\end{figure}

The average number ${\cal N}(H)$ (\ref{nh}) has been also determined numerically for percolation 
clusters in \cite{MS}, and for $c=0$, our prediction (\ref{toc}) $\tau ^{\ast }=\frac{23}{24}=0.95833...$ 
compares very well with the 
result 
$\tau ^{\ast }=0.951\pm 0.030$, obtained for $%
10^{-5}\leq H\leq 10^{-4}$.

The dimension of the support of the measure 
$D\left( 
0\right)=\frac{4}{3} \neq D_{{\rm H}},$ where $D_{{\rm %
H}}=\frac{7}{4}$ is the Hausdorff dimension of the standard hull, i.e., the 
outer
boundary of critical percolating clusters \cite{SD}. The value $ D(0)=\frac{4}{3}$ 
corresponds to dimension of the {\it accessible external 
perimeter}. A direct derivation of its exact value is given in \cite{DAA}. The complement of the accessible perimeter in the
hull is made of deep
fjords, which do close in the scaling limit and are not probed by the harmonic measure. This is in
agreement with the instability phenomenon observed on a lattice for the hull 
dimension \cite{GA}. A striking fact is the complete identity of the multifractal 
spectrum  for percolation to the corresponding results, Eqs.(\ref{tauf}-\ref{mf}), 
{\it both} for random
walks and self-avoiding walks. Seen from outside, these three fractal simple curves
are not distinguished by the harmonic measure. in fact they are the same, 
and one of the main conclusions of this study is that {\it the external frontiers of a planar 
Brownian motion, 
or of a critical percolation cluster are identical to a critical self-avoiding walk, 
with a Hausdorff dimension $D=\frac{4}{3}$.} As we have seen, this fact is 
linked to the
presence of a single universal conformal field theory (with a vanishing
central
charge $c=0$), and to the underlying presence of quantum gravity, which structures 
the associated conformal dimensions. Note that in a recent work, Smirnov \cite{smirnov1} 
proved that critical site percolation on the triangular lattice has a 
conformally invariant scaling limit, and that the discrete cluster interfaces (hulls) converge 
to the same stochastic L\"owner evolution process as the one involved for Brownian paths, 
opening the way to a rigorous derivation of percolation exponents \cite{lawler6,smirnov2}, 
previously derived in physics \cite{dennijs,cardy,SD,DAA}.\\

\textbf{\large{\textbf{VI General Conformally Scaling Curves and Higher Multifractality}}}\\

In the next sections, we present the main description of multifractal functions
results in a universal way. For multiple simple curves, we also
define the higher multifractal spectra, depending on an arbitrary number of $\alpha$-variables.
We then proceed with their derivation from conformal field theory and quantum gravity.
The geometrical findings are described in details, including well-known cases
(Ising clusters, $Q=4$ Potts model).
Finally, some geometric duality properties
 for the external boundaries in $O(N)$ and Potts models are explained. We also make explicit the relation between
 a conformally invariant scaling curve with CFT central charge $c$ \cite{duplantier11}, and the
 stochastic L\"owner process $SLE_{\kappa}$ \cite{schramm1,lawler4,lawler5,schramm2,lawleresi}. \\

Consider a single (conformally invariant) critical random cluster, generically called ${\cal C}$. Let
$H\left( z\right) $ be the potential at exterior point $z \in {\rm {\bf  C}}$, with
Dirichlet boundary conditions
$H\left({w \in \partial \cal C}\right)=0$ on the outer (simply connected) boundary $\partial \cal C$ of $\cal C$,
(or frontier  $ {\cal F} \equiv \partial \cal C$) , and
$H(w)=1$ on a circle ``at $\infty$'', i.e., of a large radius
scaling like the average size $R$ of $ \cal C$.
From the well-known theorem due to Kakutani \cite{kakutani}, $H\left( z\right)$ is identical
to the {\it harmonic measure}, i.e,
the probability that a random walker (more precisely, a Brownian motion) launched from
$z$, escapes to $\infty$ without having hit ${\cal C}$.
The multifractal formalism \cite{mandelbrot2,hentschel,frisch,halsey}
characterizes subsets ${\partial\cal C}_{\alpha }$ of boundary sites
by a H\"{o}lder exponent $\alpha ,$ and a Hausdorff
dimension $f\left( \alpha \right) ={\rm dim}\left({\partial\cal C}_{\alpha }\right)$, such that their
potential locally scales as
\begin{equation}
H\left( z \to w\in {\partial\cal C}_{\alpha }\right) \approx \left( |z-w|/R\right) ^{\alpha },
\label{ha''}
\end{equation}
in the scaling limit $a \ll r=|z-w| \ll R,$ with $a$ the underlying lattice constant.
In  2D the {\it complex} potential $\varphi(z)$ (such that the electrostatic potential
$H(z)=\Re \varphi(z)$
and field $|{\bf E}(z)|=|\varphi'(z)|$) reads for a {\it wedge} of angle $\theta$, centered at $w$:
\begin{equation}
\varphi(z) = (z-w)^{{\pi}/{\theta}}.
\end{equation}
 By Eq. (\ref{ha''}) a
H\"older exponent $\alpha$ thus defines a local equivalent ``electrostatic'' angle
$\theta={\pi}/{\alpha},$ and the MF dimension $\hat f(\theta)$
of the boundary subset with such $\theta$ is
\begin{equation}
\hat f(\theta) = f(\alpha={\pi}/{\theta}).
\label{fchapeau}
\end{equation}
Of special interest are the moments of $H$, averaged over all
realizations of ${\cal C}$
\begin{equation}
{\cal Z}_{n}=\left\langle \sum\limits_{z\in {\partial {\cal} C(r)}}H^{n}\left( z\right)
\right\rangle ,
\label{Za}
\end{equation}
 where $\partial {\cal C}(r)$ is shifted a distance $r$ outwards from $\partial \cal C$, and where $n$ can be
 a real number. In the scaling limit, one expects these moments to scale as
\begin{equation}
{\cal Z}_{n}\approx \left( r/R\right) ^{\tau \left( n\right) },  \label{Z2a'}
\end{equation}
where the multifractal scaling exponents
$\tau \left(
n\right) $ encode generalized dimensions, $D\left( n\right)=\tau \left(
n\right) /\left( n-1\right)$,
which vary in a non-linear way with $n$\cite{mandelbrot2,hentschel,frisch,halsey};
they obey the symmetric Legendre transform
$\tau \left( n\right)
+f\left( \alpha \right) =\alpha n,$ with $n=f'\left( \alpha \right), \alpha =\tau'\left( n\right)$.
From Gauss' theorem \cite{cates et witten} $\tau (1)=0.$ As said above, because of the
ensemble average (\ref{Za}), values of $f\left( \alpha \right)$
can become negative for some domains of $\alpha $
\cite{cates,cates et witten}.

Now, we consider the specific case where the fractal set ${\cal C}$ is a (conformally invariant) 
{\it simple scaling curve}, that is,
it does not contain double points. The frontier $\partial\cal C$ is thus identical with the set itself:
\begin{equation}
\partial\cal C=\cal C.
\end{equation}

Each point of the curve can then be reached from
infinity, and we can address the more refined question 
of the simultaneous behavior of the potential on both sides of the curve.
Specifically, the potential $H$ scales as
\begin{equation}
H_{+}\left( z \to w^{+}\in {\partial\cal C}_{\alpha,\alpha' }\right) \approx |z-w|^{\alpha},
\label{ha+}
\end{equation}
when approaching $w$ on one side of the scaling curve, together with the scaling
\begin{equation}
H_{-}\left( z \to w^{-}\in {\partial\cal C}_{\alpha,\alpha' }\right) \approx |z-w|^{\alpha'},
\label{ha-}
\end{equation}
on the other side.
We can then generalize the multifractal formalism to
characterize subsets ${\cal C}_{\alpha,\alpha' }$ of boundary sites $w$
by two H\"{o}lder exponents $\alpha ,\alpha'$ such that the potential 
near $w$ locally scales on the two sides of ${\cal C}$
as in 
Eqs. (\ref{ha+}) and  (\ref{ha-}). This subset is characterized by a Hausdorff 
dimension $f_2\left( \alpha,\alpha' \right) ={\rm dim}\left({\cal C}_{\alpha,\alpha' }\right)$.
The standard multifractal spectrum $f(\alpha)$ is then recovered as the supremum:
\begin{equation}
f(\alpha)={\rm sup_{\alpha'}}f_2\left( \alpha,\alpha' \right).
\label{sup}
\end{equation}
As above, one can also define two equivalent ``electrostatic'' angles from the
H\"older exponents $\alpha,\alpha'$, as
$\theta={\pi}/{\alpha},\theta'={\pi}/{\alpha'}$ and the MF dimension $\hat f_2(\theta,\theta')$ 
of the boundary subset with such $\theta,\theta'$ is then
\begin{equation}
\hat f_2(\theta,\theta') = f_2(\alpha={\pi}/{\theta},\alpha'={\pi}/{\theta'}). 
\label{fchapeau'}
\end{equation}
Define the harmonic measure
${\rm H}\left( w\right) $ as the probability that a random walker (RW) launched 
from 
infinity, {\it first} hits the  frontier ${\cal 
C}$ at point $w \in {\cal C}$. A covering 
of $\cal C$ by balls ${\cal B}(w, r)$ of radius $r$ is centered at points $w \in 
{\cal C}/\{r\}$ forming a discrete subset ${\cal C}/\{r\}$ of $\cal C$. Let ${\rm H}({\cal C} \cap {\cal B}(w, r))$
 be the 
harmonic measure of the intersection of $\cal C$ and the ball ${\cal B}(w, r)$. 
The double multifractal spectrum will be computed from the double moments of the harmonic measure 
on {\it both} sides of the random fractal curve. Let us define:
\begin{equation}
{\cal Z}_{n,n'}=\left\langle \sum\limits_{w\in {\cal C}/\{r\}}\left[{\rm H}_{+}(w)\right]^{n}
\left[{\rm H}_{-}(w)\right]^{n'}\right\rangle,
\label{ZZ''}
\end{equation}
where ${\rm H}_{+}(w) \equiv {\rm H}_{+}\left({\cal C} \cap {\cal B}(w,
r)\right)$ and ${\rm H}_{-}(w)\equiv {\rm H}_{-}\left({\cal C} \cap {\cal B}(w,r)\right)$ are respectively
the harmonic measures on the ``left'' or ``right'' sides of the random fractal. These double moments have a
multifractal scaling behavior
 \begin{equation}
{\cal Z}_{n,n'}\approx \left( r/R\right) ^{\tau_2 \left( n,n'\right) },  \label{ZZ2''}
\end{equation}
where the exponent $\tau_2 \left(n,n'\right)$ now depends on two moment orders $n,n'$.
The generalization of the usual Legendre transform of multifractal formalism Eq. (\ref{alpha}) reads
\begin{eqnarray}
\alpha &=&\frac{\partial\tau_2 }{\partial n}\left( n,n'\right) ,
\quad \alpha' =\frac{\partial\tau_2 }{\partial n'}\left( n,n'\right), \nonumber \\
f_2\left( \alpha, \alpha' \right) &=&\alpha n+\alpha' n'-\tau_2 \left( n,n'\right), \nonumber \\
n&=&\frac{\partial f_2}{\partial \alpha}\left( \alpha, \alpha'
\right) , \quad n'=\frac{\partial f_2}{\partial \alpha' }\left( \alpha, \alpha'
\right).
\label{alpha'''}
\end{eqnarray}
From definition (\ref{ZZ''}) and Eq. (\ref{ZZ2''}), we recover the one-sided multifractal spectrum as
\begin{equation}
\tau \left( n\right)=\tau_2 \left( n,n'=0\right).
\end{equation}
Putting the value $n'=0$ in the Legendre transform Eq. (\ref{alpha'''}), we obtain the identity (\ref{sup}), as it must.\\

More generally, one can consider a star configuration ${\cal S}_m$ of a number $m$, $m\geq 2$,
 of {\it similar simple scaling paths},
all originating at the same vertex $w$. The higher moments ${\cal Z}_{n_1,n_2,...,n_m}$
can then be defined
as
\begin{equation}
{\cal Z}_{n_1,n_2,...,n_m}=\left\langle \sum\limits_{w\in {\cal S}_m}\left[{\rm H}_{1}(w)\right]^{n_1}
\left[{\rm H}_{2}(w)\right]^{n_2}\cdots\left[{\rm H}_{m}(w)\right]^{n_m} \right\rangle,
\label{Z_m}
\end{equation}
where $${\rm H}_{i}(w)\equiv {\rm H}_{i}\left({\cal C} \cap {\cal B}(w,
r)\right)$$ is the harmonic measure (or, equivalently, local potential at distance $r$) in the
$i$th sector of radius
located between paths $i$ and $i+1$, with $i=1,\cdots,m$, and by periodicity $m+1 \equiv 1$.
These higher moments have a
multifractal scaling behavior
 \begin{equation}
{\cal Z}_{n_1,n_2,...,n_m}\approx \left( r/R\right) ^{\tau_m \left(n_1,n_2,...,n_m \right) },
\label{Z_m'}
\end{equation}
where the exponent $\tau_m \left(n_1,n_2,...,n_m\right)$ now depends on the set of moment orders
$n_1,n_2,...,n_m$.
The generalization of the usual Legendre transform of multifractal formalism Eq. (\ref{alpha}) now
involves a higher multifractal function $f_m\left(\alpha_1, \alpha_2,\cdots, \alpha_m \right)$,
 depending on $m$ local exponents $\alpha_i$:
\begin{eqnarray}
\alpha_i &=&\frac{\partial\tau_m }{\partial n_i}\left( \{n_i\}\right) ,
 \nonumber \\
f_m\left(\{\alpha_i\}\right) &=&\sum_{i=1}^{m}\alpha_i n_i-
\tau_m \left( \{n_i\}\right), \nonumber \\
n_{i}&=&\frac{\partial f_m}{\partial \alpha_i}\left(\{\alpha_j\}\right).
\label{alpha_m}
\end{eqnarray}

At this point, a caveat is in order. The reader could wonder about the meaning of the sum over point
$w$ in (\ref{Z_m}), if there is only
one such $m$-vertex in a star! This formal notation is kept in continuity with the $m=2$ case, and annonces the fact that exponents
$\tau_m \left(n_1,n_2,...,n_m\right)$ are calculated with inclusion of the
 Hausdorff dimension $D_m$,
associated with the star center, which reads in these notations $D_m=-\tau_m \left(0,0,...,0\right)
={\rm sup}_{\{\alpha_i\} }f_m\left(\{\alpha_i\}\right)$, and becomes negative for $m$ high enough (see section VIII below). One can define shifted
exponents $\tilde \tau_m\equiv \tau_m +D_m=\tau_m \left(n_1,n_2,...,n_m\right)-\tau_m \left(0,0,...,0\right)$,
which correspond to a different normalization, and describe the scaling of local averages
\begin{equation}
\left\langle \left[{\rm H}_{1}(w)\right]^{n_1}
\left[{\rm H}_{2}(w)\right]^{n_2}\cdots\left[{\rm H}_{m}(w)\right]^{n_m} \right\rangle \approx
\left( r/R\right) ^{\tilde \tau_m \left(n_1,n_2,...,n_m \right)}\,.
\label{H_m}
\end{equation}
By Legendre transform (\ref{alpha_m}) these exponents give the {\it subtracted} spectrum
$f_m\left(\{\alpha_i\}\right)-{\rm sup}_{\{\alpha_i\} }f_m\left(\{\alpha_i\}\right)$ directly. The latter has an
immediate physical meaning:
 the probability
$P(\{\alpha_i\})$ to find a set
of local singularity exponents $\{\alpha_i\}$ in the $m$ sectors of
an $m$-arm star scales as:
\begin{equation}
P_m(\{\alpha_i\})\propto R^{f_m\left(\{ {\alpha}_{i}\}\right)}/R^{\,{\rm sup} f_m} \, .
\label{probastar}
\end{equation}

From definition (\ref{Z_m}) and Eq. (\ref{Z_m'}), we get the lower
$(m-1)$-multifractal spectrum as
\begin{equation}
\tau_m^{[m-1]} \left(n_1,n_2,\cdots, n_{m-1}\right)=\tau_m \left( n_1,n_2,\cdots, n_{m-1}, n_m=0\right).
\end{equation}
In these exponents, the subscript $m$ stays unchanged since it counts the number of arms
of the star, while the potential is evaluated only at $m-1$ sectors among the $m$ possible.
More generally, one can define exponents
$$\tau_m^{[p]} \left(n_1,n_2,\cdots, n_{p}\right)=
\tau_m \left( n_1,n_2,\cdots, n_{p}~; n_{p-1}=0,\cdots, n_m=0\right),$$
where $p$ takes any value in $1\leq p \leq m$. Note that according to the commutativity of the
star algebra for exponents between mutually-avoiding paths
(see Eq.(\ref{x}) and below), the result does not depend on the
choice of the $p$ sectors among $m$. Putting the value $n_m=0$ in the Legendre transform Eq. (\ref{alpha_m}), we obtain
the identity:\\
\begin{equation}
f_m^{[m-1]}(\alpha_1,\cdots,\alpha_{m-1})={\rm sup_{\alpha_m}}
f_m\left( \alpha_1, \alpha_2,\cdots, \alpha_m \right).
\label{sup_m}
\end{equation}
Note that the usual $f(\alpha)$ spectrum is in these notations $f_2^{[1]}(\alpha)$.\\

 \textbf{\large{\textbf{VII Conformal Invariance and Quantum Gravity}}}\\

 Let us
 now follow the main lines of the derivation of exponents $\tau\left(
n\right) $, hence $f(\alpha),$ or, more generally, $\tau_m \left(n_1,n_2,...,n_m\right)$ and
 $f_m\left( \alpha_1, \alpha_2,\cdots, \alpha_m \right)$ by generalized {\it conformal invariance}.
By definition of the ${\rm H}$-measure, $n$ {\it independent} RW's, or Brownian paths ${\cal B}$
in the scaling limit, starting at the same
point a distance $r$ away
from the cluster's hull's frontier $\partial \cal C$, and diffusing
without hitting $\partial \cal C$, give a geometric representation of the $n^{th}$
moment$, H^{n},$ in Eq.(\ref{Z}) for $n$ {\it integer}. Convexity yields analytic
continuation for arbitrary $n$'s. Let us introduce the notation $A\wedge B$ for
two random sets conditioned to traverse, {\it without mutual intersection}, the
annulus ${\cal%
D}\left( r, R\right) $ from the inner boundary circle of radius $r$ to the outer
one at distance $R$,
and{\it \ }$A\vee B$ for two {\it independent}, thus
possibly
intersecting, sets \cite{duplantier8}. With this notation, the ``probability'' (actually the associated
grand canonical partition function) that the Brownian paths
and cluster are in a configuration ${\partial \cal C}\wedge
\left( \vee {\cal B}\right) ^{n}\equiv {\partial\cal C}\wedge {n}$, is expected to scale for $%
R/r\rightarrow \infty $ as
\begin{equation}
{\cal P}_{R}\left( {\partial\cal C}\wedge n\right) \approx
\left( r/R\right) ^{x\left( n\right) },
\label{xp}
\end{equation}
where the scaling exponent $x\left(n\right)$ depends on $n.$
In terms of definition (\ref{xp}), the harmonic measure moments (\ref{Za}) simply
scale as
${\cal Z}_{n}\approx R^2{\cal P}_{R}\left( {\partial \cal C}\wedge n\right)$
\cite{cates et witten,duplantier8},
 which, combined with Eq. (\ref{Z2}), leads to
\begin{equation}
\tau \left( n\right) =x\left( n\right) -2.  \label{tt'}
\end{equation}

To calculate these exponents, we use the fundamental mapping of the conformal field theory in the {\it
plane} $%
{\rm \bf R}^{2},$ describing a critical statistical system, to the
CFT on a fluctuating abstract random Riemann surface, i.e., in presence of {\it quantum gravity}
\cite{polyakov,david2,DK}. Two universal functions $U,$ and $V,$ which now depend on the central charge $c$ of
the CFT, describe this map:
\begin{eqnarray}
U\left( x\right) &=&x\frac{x-\gamma}{1-\gamma} , \hskip2mm V\left( x\right)
=\frac{1}{4}\frac{x^{2}-\gamma^2}{1-\gamma},  \label{Ua}
\end{eqnarray}
with
\begin{equation}
V\left( x\right) \equiv U\left(
{\textstyle{1 \over 2}}%
\left( x+\gamma%
\right) \right).
\end{equation}
The parameter $\gamma$ is the {\it string susceptibility exponent}
of the random 2D surface (of genus zero),
bearing the CFT of central charge $c$\cite{polyakov}; $\gamma$ is the solution of
\begin{equation}
c=1-6{\gamma}^2(1-\gamma)^{-1}, \gamma \leq 0.
\label{cgamma}
\end{equation}
For two arbitrary\ random sets $A,B,$
with boundary scaling exponents in the {\it half-plane} $\tilde{x}\left( A\right) ,\tilde{x}\left(
B\right),$ the scaling exponent $x\left( A\wedge
B\right)$, as in (\ref{xp}), has the universal structure \cite{duplantier8,duplantier9,duplantier11}
\begin{eqnarray}
x\left( A\wedge B\right) &=&2V\left[ U^{-1}\left( \tilde{x}\left( A\right)
\right) +U^{-1}\left( \tilde{x}\left( B\right) \right) \right],  
\label{xa}
\end{eqnarray}
where $U^{-1}\left( x\right) $ is the {\it positive} inverse function of $U$
\begin{equation}
U^{-1}\left( x\right) =\frac{1}{2}\left(\sqrt{4(1-\gamma)x+\gamma^2}+\gamma\right) .  \label{U1a}
\end{equation}
Note that one has the shift relation
\begin{equation}
U^{-1}\left( x\right) =\frac{1}{2}V^{-1}\left( x\right)+\frac{1}{2}\gamma\ ,
\label{shift}
\end{equation}
where
\begin{equation}
V^{-1}\left( x\right) =
\sqrt{4(1-\gamma)x+\gamma^2}\ .
\label{V1}
\end{equation}  
$U^{-1}\left( \tilde{x} \right)$ is, on the random Riemann surface, the boundary
scaling dimension corresponding to $\tilde{x}$ in the half-plane  ${\rm \bf R} \times
{\rm \bf R}^{+}$, and the sum of $U^{-1}$ functions in Eq. (\ref{xa})
is a {\it linear} representation of the product of two
``boundary operators'' on the random surface, as the condition $A \wedge B$ for two
mutually-avoiding sets is purely {\it topological} there. The sum is mapped back
by the function $V$ into the scaling dimensions in ${\rm \bf R}^2$\cite{duplantier11}.

For the harmonic exponents $x(n) \equiv x\left({\partial\cal
C}\wedge n \right)$ in (\ref{xp}), we use (\ref{xa}). The {\it external
boundary} exponent $\tilde{x}\left({\partial \cal C}\right)$
obeys
\begin{equation}
U^{-1}\left( \tilde{x}\right) =1-\gamma,
\label{xtilde}
\end{equation}
which we derive either directly,
or from Makarov's theorem:
\begin{equation}
\tau'(n=1)=\frac{ dx}{ d n}(n=1)=1.
\end{equation}
The bunch of $n$
independent Brownian paths have simply
$\tilde{x}\left( \left( \vee {\cal B}\right) ^{n}\right)=n,$
since $\tilde{x}%
\left( {\cal B}\right)=1$ \cite{duplantier8}. Thus we obtain
\begin{equation}
x\left( n\right) =2V\left(
1-\gamma
+U^{-1}\left( n\right) \right).  \label{finaa}
\end{equation}
This finally gives from (\ref{Ua})(\ref{U1a}) $\tau(n)=x(n)-2$\cite{duplantier11}:
\begin{equation}
\tau\left( n\right) =\frac{1}{2}(n-1)+\frac{1}{4}\frac{2-\gamma}{1-\gamma}
[\sqrt{4(1-\gamma)n+{\gamma}^2}-(2-\gamma)]\ .
\label{tauoriginal}
\end{equation}

 Similar exponents
associated with moments later appeared in the context of the $SLE$ process
(see II in \cite{lawler4},\cite{lawleresi}) ; see also \cite{hastings} for Laplacian random walks.\\

The Legendre transform is easily performed to yield:
\begin{eqnarray}
\alpha &=&\frac{d{\tau} }{dn}\left( n\right)=\frac{1}{2} +\frac{1}{2}
\frac{2-\gamma}{\sqrt{4(1-\gamma)n+{\gamma}^2}};
\label{a'}
\\
\nonumber
\\
f\left( \alpha \right)&=& \frac{1}{8}\frac{(2-\gamma)^2}{1-\gamma}\left(3- \frac{1}{2\alpha
-1}\right)
-\frac{1}{4}\frac{\gamma^2}{1-\gamma}\alpha,
\label{foriginal}
\\
\quad \alpha &\in& \left(
{\textstyle{1 \over 2}}%
,+\infty \right) .
\nonumber
\end{eqnarray}
Using the identities in terms of central charge $c$:
\begin{eqnarray}
\label{c,gamma}
\frac{1}{4}\frac{(2-\gamma)^2}{1-\gamma}&=&\frac{25-c}{24}\\
\nonumber
\frac{1}{4}\frac{\gamma^2}{1-\gamma}&=&\frac{1-c}{24},
\end{eqnarray}
we find
\begin{eqnarray}
\tau\left( n\right) &=&\frac{1}{2}(n-1)+\frac{25-c}{24}
\left(\sqrt{\frac{24n+1-c}{25-c}}-1\right)
\\
\quad n&\in& \left[ n^{\ast}=
-\frac{1-c}{24}
,+\infty \right) .
\nonumber
\\
\alpha &=&\frac{d{\tau} }{dn}\left( n\right)=\frac{1}{2} +\frac{1}{2}
\sqrt{\frac{25-c}{24n+1-c}};
\label{a'bis}
\nonumber
\\
f\left( \alpha \right)&=& \frac{25-c}{48}\left(3- \frac{1}{2\alpha
-1}\right)
-\frac{1-c}{24}\alpha,
\label{foriginalbis}
\\
\quad \alpha &\in& \left(
{\textstyle{1 \over 2}}%
,+\infty \right).
\nonumber
\end{eqnarray}
This formalism immediately allows generalizations. For instance, in place
of $n$ random walks, one can consider a set of $n$ {\it independent
self-avoiding} walks $\cal P$, which avoid the cluster fractal boundary, except
for their common anchoring point. The associated multifractal exponents $
x\left( {\partial\cal C}\wedge  \left( \vee {\cal P}\right)^{n} \right)$ are
given by (\ref{finaa}), with the argument $n$ in $U^{-1}(n)$ simply
replaced by  ${\tilde x}\left( \left( \vee {\cal P}\right) ^{n}\right) =n{\tilde x}
\left( {\cal P}\right) =\frac{5}{8}n $ \cite{duplantier8}. These exponents govern
the universal multifractal behavior of the moments
of the probability that a SAW escapes from $\cal C$. One then gets a spectrum $\bar f$ such that
${\bar f}\left(\bar\alpha=\tilde{x} \left( {\cal P}\right)\pi/\theta \right)
= f\left(\alpha=\pi/\theta\right)={\hat f}(\theta)$, thus unveiling the {\it same invariant}
underlying wedge distribution as the harmonic measure, (see also \cite{cardy2}).\\

\textbf{\large{\textbf{VIII Higher Multifractal Spectra}}}\\

In analogy to Eqs. (\ref{tt'}), (\ref{finaa}), the exponent $\tau_2(n,n')$ is associated with a scaling dimension
$x_2(n,n')$ by 
\begin{eqnarray}
\nonumber
\tau_2(n,n')&=&x_2(n,n')-2 \\
x_2(n,n')&=&2V\left[
1-\gamma
+U^{-1}\left( n\right)+U^{-1}\left( n'\right) \right].  \label{finaa'}
\end{eqnarray}

The calculation of the double Legendre transform Eq. (\ref{alpha'''}) is as follows. We start with the notation 
for the total quantum gravity scaling dimension:
\begin{equation}
\delta\equiv 1-\gamma
+U^{-1}\left( n\right)+U^{-1}\left( n'\right)\ .
\label{delta-n,n'}
\end{equation}
This gives explicitly:
\begin{equation}
\delta= 1
+\frac{1}{2}\sqrt{4(1-\gamma)n+\gamma^2}+\frac{1}{2}\sqrt{4(1-\gamma)n'+\gamma^2}\ .
\label{x'}
\end{equation}
Then, we have 
\begin{eqnarray}
\alpha &=&\frac{\partial x_2 }{\partial n}\left( n,n'\right)=2V'(\delta)\,\frac{\partial \delta }{\partial n}
\end{eqnarray}
and since
$$V'(x)=\frac{1}{2}\frac{x}{1-\gamma}\ ,$$
we finally get 
\begin{eqnarray}
\alpha=\frac{\delta}{\sqrt{4(1-\gamma)n+\gamma^2}} \ ,\, \alpha'=\frac{\delta}{\sqrt{4(1-\gamma)n'+\gamma^2}}\ .
\label{alpha-n}
\end{eqnarray}
 A useful consequence is the identity
\begin{equation}
\delta={\left[1-\frac{1}{2}\left(\frac{1}{\alpha}+\frac{1}{\alpha'}\right)\right]}^{-1}\ .
\label{identity}
\end{equation}
Equation (\ref{alpha-n}) can be inverted into
\begin{equation}
n=\frac{1}{4(1-\gamma)}\left(\frac{\delta^2}{\alpha^2}-\gamma^2\right)=V\left(\frac{\delta}{\alpha}\right)\ ,
\label{n}
\end{equation}
where use was made of (\ref{Ua}) for $V$.
This allows the simple expression of $f_2$
\begin{equation}
f_2\left( \alpha, \alpha' \right)=2-V(\delta)+\alpha V\left(\frac{\delta}{\alpha}\right)+
\alpha' V\left(\frac{\delta}{\alpha'}\right)\ .
\label{f_2}
\end{equation}
Reordering the $\delta$ terms with use of (\ref{Ua}), and recalling identity (\ref{identity}) for $\delta$,
 finally gives after some calculations the explicit formulae
\begin{eqnarray}
f_2\left( \alpha, \alpha' \right)&=&\frac{25-c}{12}-\frac{1}{2(1-\gamma)}
{\left[1-\frac{1}{2}\left(\frac{1}{\alpha}+\frac{1}{\alpha'}\right)\right]}^{-1} \nonumber \\
& &-\frac{1-c}{24}\left(\alpha+\alpha'\right),
\label{f_2c}
\end{eqnarray}
\begin{equation}
{\alpha}=\frac{1}{\sqrt{4(1-\gamma)n+\gamma^2}}\left[1
+\frac{1}{2}\left(\sqrt{4(1-\gamma)n+\gamma^2}+\sqrt{4(1-\gamma)n'+\gamma^2}\right)\right],
\label{alphann'c}
\end{equation}
where the central charge $c$ and the parameter $\gamma$ are related by Eqs. (\ref{c,gamma}). 
This doubly multifractal spectrum possesses the requested properties, like
${\rm sup}_{\alpha'} f_2(\alpha, \alpha')=f(\alpha),$ where $f(\alpha)$ is (\ref{foriginal}) above.

This double multifractality can be generalized to higher ones by considering {\it star configurations} made of 
$m$ simple scaling paths all originating at the same vertex, as in (\ref{Z_m}), with the following 
{\it poly-multifractal} results. The $m$-order case will be given by
\begin{eqnarray}
\nonumber
\tau_m \left(n_1,n_2,...,n_m\right)&=&x_m \left(n_1,n_2,...,n_m\right)-2 \\
x_m \left(n_1,n_2,...,n_m\right)&=&2V\left[
\tilde \Delta_m
+U^{-1}\left( n_1\right)+U^{-1}\left( n_2\right)+...+U^{-1}\left( n_m\right) \right].  \label{x_m}
\end{eqnarray}
Here $\tilde \Delta_m$ is the quantum gravity boundary scaling dimension of the $m$-star ${\cal S}_m$ made of 
$m$ ( simple) scaling paths. 
According to the star algebra we have:
\begin{equation}
\tilde \Delta_m=m\, U^{-1}\left( \tilde x_1\right)=\frac{m}{2} U^{-1}\left( \tilde{x}_2\right)=m \frac{1-\gamma}{2}\ ,
\label{Delta_m}
\end{equation}
whre $\tilde {x}_2\equiv\tilde x$ is the boundary scaling dimension of a scaling 
path, i.e., a $2$-star, already considered in Eq. (\ref{xtilde}), and such that $U^{-1}\left( \tilde{x}\right)=1-\gamma$.
We therefore arrive at a total (boundary) quantum scaling dimension
\begin{equation}
\delta_m=m \frac{1-\gamma}{2}
+\sum_{i=1}^{m}U^{-1}\left( n_i\right)\ ,
  \label{delta_m}
\end{equation}
such that 
\begin{equation}
x_m \left(n_1,n_2,...,n_m\right)=2V\left(\delta_m\right)\ .
\label{x-delta_m}
\end{equation} 
Using the shift identity (see (\ref{shift})) 
$$U^{-1}\left( n\right)=\frac{\gamma}{2}+\frac{1}{2}V^{-1}\left( n \right),$$ where we recall that
$$V^{-1}\left( n \right)=\sqrt{4(1-\gamma)n+\gamma^2},$$ we also have
\begin{equation}
\delta_m= \frac{m}{2}
+\frac{1}{2}\sum_{i=1}^{m}V^{-1}\left( n_i\right)\ .  
\label{delta_m'}
\end{equation}
The multiple Legendre transform (\ref{alpha_m}) is performed as above for the case $m=2$. 
We have 
\begin{eqnarray}
\alpha_i &=&\frac{\partial x_m }{\partial n_i}\left( \{n_j\}\right)=2V'(\delta_m)\,\frac{\partial \delta_m }{\partial n_i}
=V'(\delta_m)\, \left[V^{-1}\left( n_i\right)\right]'\ ,
\end{eqnarray}
so we get 
\begin{eqnarray}
\alpha_i=\frac{\delta_m}{\sqrt{4(1-\gamma)n_i+\gamma^2}}=\frac{\delta_m}{V^{-1}\left( n_i\right)} \ ,
\label{alpha-n_i}
\end{eqnarray}
or, equivalently
\begin{eqnarray}
V^{-1}\left( n_i\right)=\frac{\delta_m}{\alpha_i} \ ,
\label{alpha-n_i'}
\end{eqnarray}
inverted into
\begin{equation}
n_i=V\left(\frac{\delta_m}{\alpha_i}\right)\ .
\label{n_i}
\end{equation}
 One gets from Eqs. (\ref{delta_m'}) and (\ref{alpha-n_i'})
\begin{equation}
\delta_m=\left(1-\frac{1}{2}\sum_{i=1}^m {\alpha_i}^{-1}\right)^{-1}\ .
\label{identity'}
\end{equation}
This allows the simple expression of $f_m$
\begin{equation}
f_m\left( \{\alpha_i\}\right)=2-V(\delta_m)+\sum_{i=1}^m\alpha_i V\left(\frac{\delta_m}{\alpha_i}\right)\ .
\label{f_m}
\end{equation}
Reordering the $\delta_m$ terms with use of (\ref{Ua}), and recalling identity (\ref{identity'}) for $\delta_m$,
 finally gives after some calculations the explicit formulae
\begin{eqnarray}
f_m\left(\{ {\alpha}_{i=1,...,m}\}\right)&=&2+\frac{\gamma^2}{2(1-\gamma)}-
\frac{1}{8(1-\gamma)}{m}^2{\left(1-\frac{1}{2}\sum_{i=1}^{m}{\alpha}_{i}^{-1}\right)}^{-1} \nonumber \\
& &-\frac{\gamma^2}{4(1-\gamma)}\sum_{i=1}^{m}{\alpha}_{i}\ ,
\label{faig}
\end{eqnarray}
with
\begin{equation}
{\alpha}_{i}=\frac{1}{\sqrt{4(1-\gamma)n_{i}+\gamma^2}}\left({\frac{m}{2}}+
\frac{1}{2}\sum_{j=1}^{m}\sqrt{4(1-\gamma)n_{j}+\gamma^2}\right)\ .\label{aic}
\end{equation}
Substituting expressions (\ref{c,gamma}) gives in terms of $c$
\begin{eqnarray}
f_m\left(\{ {\alpha}_{i=1,...,m}\}\right)&=&\frac{25-c}{12}-
\frac{1}{8(1-\gamma)}{m}^2{\left(1-\frac{1}{2}\sum_{i=1}^{m}{\alpha}_{i}^{-1}\right)}^{-1} \nonumber \\
& &-\frac{1-c}{24}\sum_{i=1}^{m}{\alpha}_{i} \ .
\label{faic}
\end{eqnarray}
The domain of definition of the poly-multifractal function $f$ is independent of $c$ and given by
\begin{equation}
1-\frac{1}{2}\sum_{i=1}^{m}{\alpha}_{i}^{-1} \geq 0,
\label{domain}
\end{equation}
as verified by Eq. (\ref{aic}). The two-sided case (\ref{f_2c}) above is recovered for $m=2$, while the self-avoiding walk case (\ref{fai}) is recovered for
$\gamma=-1/2$, $c=0$. Notice that the case $f_{m=1}(\alpha_1)$ corresponds to the potential in the vicinity of the tip
of a conformally invariant scaling path, and differs from the usual
$f(\alpha)={\rm sup}_{\alpha'} f_2(\alpha,\alpha'))$ spectrum, which describes
the potential on one side of the scaling path.

We can also substitute equivalent ``electrostatic'' angles $\theta_i=\pi/\alpha_i$ to variables $\alpha_i$.
This gives a new distribution:
\begin{eqnarray}
{\hat f}_m\left(\{ {\theta}_{i=1,...,m}\}\right)\equiv f_m\left(\{ {\alpha}_{i=1,...,m}\}\right)
&=&2+\frac{\gamma^2}{2(1-\gamma)}-\frac{\gamma^2}{4(1-\gamma)}\sum_{i=1}^{m}\frac{\pi}{\theta}_{i} \nonumber \\
& &-
\frac{1}{8(1-\gamma)}{m}^2{\left(1-\frac{1}{2\pi}\sum_{i=1}^{m}{\theta}_{i}\right)}^{-1}\ .
\label{fthetaig}
\end{eqnarray}
The domain of definition of distribution $\hat f_m$ is the image of domain (\ref{domain}) in $\theta$-variables:
 \begin{equation}
\sum_{i=1}^{m}{\theta}_{i} \leq 2\pi\ .
\label{domaint}
\end{equation}
The total {\it electrostatic} angle is thus less than $2\pi$, which simply accounts for the electrostatic screening of local 
wedges by fractal randomness, as expected.

The maximum of $f_m$ or ${\hat f}_m$ is by construction obtained for $n_i=0,\ \forall i=1,...,m.$
Eq. (\ref{aic}) gives the values of singularity exponents $\hat \alpha_i$ at the maximum of $f_m$:
\begin{equation}
\hat \alpha_i=\frac{\pi}{\hat\theta_i}=\frac{m}{2}\left(1-\frac{1}{\gamma}\right)^{-1}, \ \forall i=1,...,m\ ,
\label{hataic}
\end{equation}
corresponding to a maximum value of $f_m$ of $\hat f_m$:
\begin{eqnarray}
\nonumber
{\rm sup}f_m= f_m\left(\{ {\hat \alpha}_{i=1,...,m}\}\right)
={\hat f}_m\left(\{ {\hat \theta}_{i=1,...,m}\}\right)&=&
2-2V(\tilde \Delta_m)\\&=&2+\frac{\gamma^2}{2(1-\gamma)}-
\frac{1}{8(1-\gamma)}{m}^2 .
\label{supf}
\end{eqnarray}
The interpretation of the poly-multifractal spectrum can be understood as follows. The probability 
$P(\{\alpha_i\})\equiv\hat P(\{\theta_i\})$ to find a set 
of local singularity exponents $\{\alpha_i\}$ or equivalent angles $\{\theta_i\}$ in the $m$ sectors of
an $m$-arm star is given by the ratio
\begin{equation}
P_m(\{\alpha_i\})\propto R^{f_m\left(\{ {\alpha}_{i}\}\right)}/R^{\,{\rm sup} f_m}
\label{proba_m} 
\end{equation}
of the respective number of configurations to the total one. 
We therefore arrive at a probability, here written in terms of the equivalent electrostatic angles:
\begin{eqnarray}
\hat P_m(\{\theta_i\})&\propto&R^{{\hat f}_m\left(\{ {\theta}_{i}\}\right)-{\hat f}_m(\{ {\hat \theta}_{i}\})}\ ,\\
\hat f_m\left(\{ {\theta}_{i}\}\right)- \hat f_m(\{ \hat {\theta}_{i}\})&=&
-\frac{1}{8(1-\gamma)}{m}^2{\left(\frac{2\pi}{\sum_{i=1}^{m}{\theta}_{i}}-1\right)}^{-1} \nonumber \\
& &-\frac{\gamma^2}{4(1-\gamma)}\sum_{i=1}^{m}\frac{\pi}{\theta}_{i}\ .
\label{ftheta-supf}
\end{eqnarray}  
For a large scaling star, the dominant set of singularity exponents
$\{\hat \alpha_i\}$, or wedge angles $\{\hat \theta_i\}$, is thus given by the symmetric set of values (\ref{hataic}).\\

\textbf{\large{\textbf{IX Analysis of Multifractal Dimensions and Spectra}}}\\

Let us collect the results for the one-sided functions $\tau(n)$, $D(n)$, and $f(\alpha)$.
Each conformally invariant random system
 is labelled by its {\it central charge} $c$, $c\leq 1$.
 The multifractal exponents $\tau(n)$ and generalized dimensions $D(n)$ of a simply
 connected CI boundary then read explicitly:
\begin{eqnarray}
\tau\left( n\right) &=&\frac{1}{2}(n-1)+\frac{25-c}{24}
\left(\sqrt{\frac{24n+1-c}{25-c}}-1\right),
\\
\nonumber
\\
D\left( n\right) &=&\frac{\tau\left( n\right)}{n-1}=\frac{1}{2}+
{\left(\sqrt{\frac{24n+1-c}{25-c}}+1\right)}^{-1}, 
\label{D''}\\
\quad n&\in& \left[ n^{\ast}=
-\frac{1-c}{24}
,+\infty \right)\ ;
\nonumber
\end{eqnarray}
with a Legendre transform~:
\begin{eqnarray}
\alpha &=&\frac{d{\tau} }{dn}\left( n\right)=\frac{1}{2} +\frac{1}{2}
\sqrt{\frac{25-c}{24n+1-c}};
\label{a'a}
\\
\nonumber
\\
\label{f''}
f\left( \alpha \right)&=& \frac{25-c}{48}\left(3- \frac{1}{2\alpha
-1}\right)
-\frac{1-c}{24}\alpha,
\\
\quad \alpha &\in& \left(
{\textstyle{1 \over 2}}%
,+\infty \right) .
\nonumber
\end{eqnarray}
It is interesting to note that the general multifractal function (\ref{f''})
possesses the invariance property (\ref{inv}), since it also reads
\begin{eqnarray}
f\left( \alpha \right)-\alpha&=& \frac{25-c}{24}
\left[1-\frac{1}{2}\left(2\alpha -1 + \frac{1}{2\alpha -1}\right)\right].
\label{f-aa}
\end{eqnarray}

Notice that the generalized dimensions $D(n)$ satisfy, for any
$c$, $\tau'(n=1)=D(n=1)=1$, or
equivalently $f(\alpha=1)=1$, i.e., {\it Makarov's theorem} \cite{makarov},
valid for any simply connected boundary curve. From (\ref{D''},\ref{a'a}) we also remark a fundamental relation,
independent of $c$:
\begin{equation}
3-2D(n)=1/\alpha=\theta/\pi.
\label{Dtheta}
\end{equation}
We also have the {\it superuniversal} bounds: $\forall c, \forall n,\frac{1}{2}=D(\infty) \leq D(n)
\leq D(n^{\ast})=\frac{3}{2}$,
hence $0 \leq \theta\leq 2\pi$.
We arrive at the geometrical multifractal distribution
of wedges $\theta$ along the boundary:
\begin{eqnarray}
\hat f(\theta)=f\left(\frac{\pi}{\theta}\right)=\frac{\pi}{\theta}-\frac{25-c}{12}
 \frac{(\pi-\theta)^2}{\theta (2\pi -\theta)}\ .
\label{fchap}
\end{eqnarray}
Remarkably enough, the second term also describes the contribution by a wedge to the
density of electromagnetic modes in a cavity \cite{BD}. The simple shift in (\ref{fchap}), $25 \to 25 -c$,
from the $c=0$ case to general values of $c$ can then be related to results of conformal
invariance in a wedge \cite{DuCa}. The partition function for the two sides of a wedge of angle $\theta$
and size $R$, in a CFT of central charge $c$, indeed scales as \cite{Ca}
\begin{equation}
 \hat {\cal Z} (\theta) \approx R^{-\frac{c}{12}\frac{(\pi-\theta)^2}{\theta (2\pi -\theta)}}\ .
\label{hatZ}
\end{equation}
Thus, one can view the $c$ dependance of result (\ref{fchap}) as a renormalization of the number of
sites with wedge angle $\theta$ along a self-avoiding scaling curve by a partition
factor $[\hat {\cal Z}(\theta)]^{-1}$,
representing the presence of a $c$-CFT along such wedges.
The maximum of $f(\alpha)$ corresponds to $n=0$, and gives the dimension $D_{\rm EP}$ of
 the support of the measure, i.e., 
the {\it accessible} or {\it external perimeter} as: 
\begin{eqnarray}
D_{\rm EP}&=&{\sup}_{\alpha}f(\alpha)=f(\alpha(n=0))\\&=&D(0)=\frac{3-2\gamma}{2(1-\gamma)} 
=\frac{3}{2}-\frac{1}{24}\sqrt{1-c}\left(\sqrt{25-c}-\sqrt{1-c}\right).
\label{D(c)}
\end{eqnarray}
This corresponds to a {\it typical exponent}
\begin{equation}
\hat\alpha={\alpha(0)}=1-\frac{1}{\gamma}=\left(\frac{1}{12}\sqrt{1-c}\left(\sqrt{25-
c}-\sqrt{1-c}\right)\right)^{-1}=(3-2D_{\rm EP})^{-1}\ .
\label{halpha}
\end{equation}
This corresponds to a {\it typical wedge angle}
\begin{equation}
\hat\theta={\pi}/{\hat \alpha}=\pi(3-2D_{\rm EP})\ .
\label{htheta} 
\end{equation}
In analogy to probability (\ref{proba_m}) for multiple angles, the probability $P(\alpha)$ to find a singularity
exponent $\alpha$ 
or, equivalently, $\hat P (\theta)$ to find an equivalent opening angle $\theta$ 
along the frontier is
\begin{equation}
P(\alpha)=\hat P(\theta)\propto R^{f(\alpha)- f(\hat\alpha)} \ .
\end{equation} 
Using the values found above, this probability can be recast as (see also \cite{cardy2})
\begin{equation}
P(\alpha)=\hat P(\theta)\propto \exp\left[-\frac{1}{24}\ln R\left(\sqrt{1-c}\sqrt{\omega}-\frac{\sqrt{25-
c}}{2 \sqrt \omega} \right)^2\right]\ ,
\label{prob}
\end{equation}
where  $$\omega=\alpha-\frac{1}{2}=\frac{\pi}{\theta}-\frac{1}{2}\ .$$
The multifractal functions $f\left( \alpha \right)-\alpha
=\hat f(\theta)-\frac{\pi}{\theta}$ 
are {\it invariant} when taken at primed variables  
such that
\begin{equation}
2\pi=\theta+{\theta}^{\prime}=\frac{\pi}{\alpha}+\frac{\pi}{{\alpha}^{\prime}}, 
\label{tetateta''}
\end{equation}
which corresponds to the complementary domain of the wedge $\theta$. 
This condition reads also $D(n)+D(n')=2.$ 
This basic symmetry, first observed and studied in \cite{BDH} for the $c=0$ result of \cite{duplantier8}, 
is valid 
for {\it any} conformally invariant boundary.

\begin{figure}[t]
\centerline{\epsfig{file=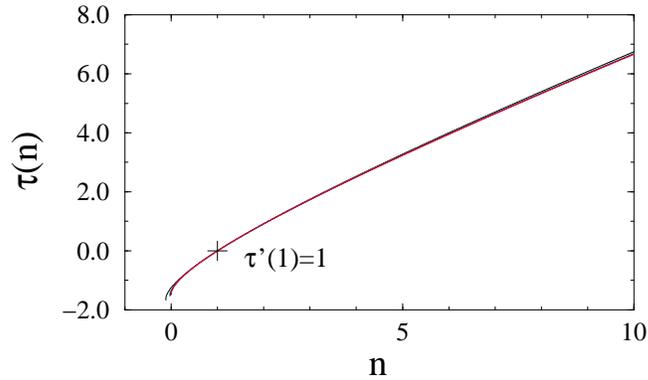,width=8.5cm}}
\caption{Universal multifractal exponents
$\tau(n)$. The curves are indexed by the central charge $c$, and correspond
respectively to the same colors as in Fig. 7 below: (black: 2D spanning trees ($c=-2$); green: self-avoiding or random walks, and 
percolation ($c=0$); blue:  
Ising clusters or $Q=2$ Potts clusters ($c=\frac{1}{2}$); red: $N=2$ loops, or $Q=4$ 
Potts clusters 
($c=1$). The curves are almost
indistinguishable at the scale shown.} 
\label{Figure4}
\end{figure}
\begin{figure}[t]
\centerline{\epsfig{file=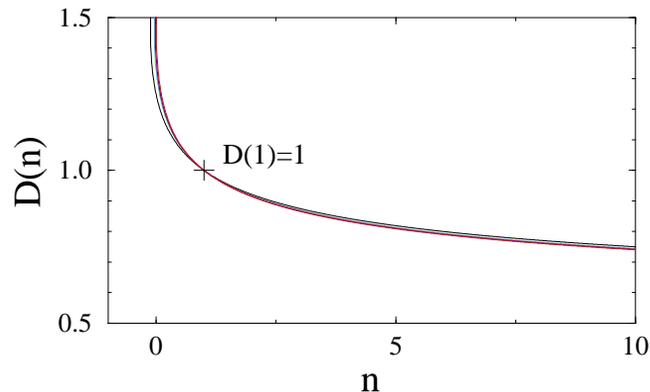,width=8.5cm}}
\caption{Universal generalized dimensions
$D(n)$. The curves are indexed by the same colors as in Fig. 7 below but are almost
indistinguishable at the scale shown.} 
\label{Figure5}
\end{figure}
The multifractal exponents $\tau(n)$ (figure 5) or generalized dimensions $D(n)$ (figure 6) appear as quite 
similar for various values of 
$c$, and a numerical simulation would hardly distinguish the different universality 
classes, while the $f(\alpha)$ functions, as we shall see, do distinguish these classes, especially for 
negative $n$, i.e. large $\alpha$.
In Fig. 7 are displayed the multifractal functions $f$, Eq. (\ref{f''}),  
corresponding to various values of 
$-2 \leq c \leq 1$, or, equivalently, to a number of components 
$N \in [0, 2]$, and $Q \in [0,4]$ in the $O(N)$ or Potts models (see below).  
\begin{figure}[t]
\centerline{\epsfig{file=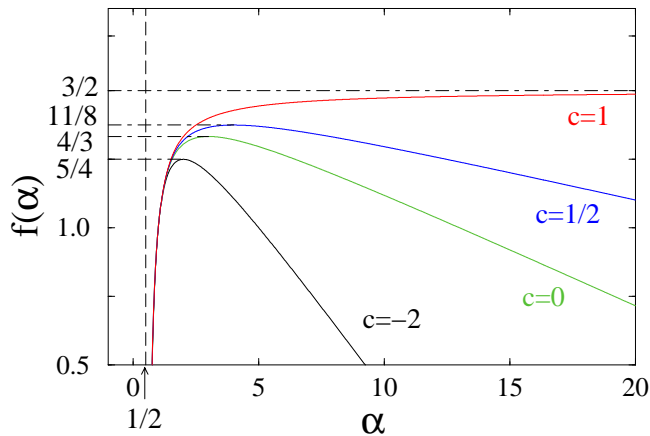,width=8.5cm}}
\caption{Universal harmonic multifractal spectra
$f(\alpha)$. The curves are indexed by the central charge $c$, and correspond
respectively to: 2D spanning trees ($c=-2$); self-avoiding or random walks, and 
percolation ($c=0$); 
Ising clusters or $Q=2$ Potts clusters ($c=\frac{1}{2}$); $N=2$ loops, or $Q=4$ 
Potts clusters 
($c=1$). The maximal dimensions are those of the {\it 
accessible} frontiers. The left branches of the various $f(\alpha)$ curves are largely
indistinguishable, while their right branches split for large $\alpha$, 
corresponding
 to negative values 
of $n$.} 
\label{Figure1}
\end{figure}  
The singularity at $\alpha=\frac{1}{2}$, or $\theta=2\pi$, in the multifractal functions $f$, or $\hat f$, 
corresponds to boundary points with a needle local geometry, and Beurling's theorem \cite{ahlfors}
indeed insures the H{\"o}lder 
exponents 
$\alpha$ to be bounded below 
by $\frac{1}{2}$. This corresponds to large values of $n$, where, asymptotically, for {\it any}
 universality class,
\begin{equation}
\forall c, \lim_{n \to \infty} D(n)=\frac{1}{2}.
\label{1/2}
\end{equation}

The right branch of $f\left( \alpha \right) $ has a linear asymptote 
\begin{equation}
\lim_{\alpha \rightarrow \infty} f\left(\alpha \right)/{\alpha} = n^{\ast}=-(1-c)/24. 
\end{equation}
The {\it limit} multifractal spectrum is obtained for $c=1$, which exhibits an {\it exact} example 
of a {\it left-sided} MF spectrum, with an asymptote 
$f\left(\alpha \to \infty, c=1\right)\to \frac{3}{2}$ (Fig. 7). It corresponds to singular boundaries where 
${\hat f}\left(\theta \to 0, c=1\right)=\frac{3}{2}=D_{\rm EP}$, i.e., where the 
external perimeter is everywhere dominated by ``{\it fjords}'', with typical angle $\hat \theta =0$.
It is tempting to call it the ``Ultimate Norway''.

The $\alpha \to \infty$ behavior corresponds to moments 
of lowest order $n\rightarrow {n^{\ast}}$, where 
$D(n)$ reaches its maximal value: $\forall c, D(n^{\ast})=\frac{3}{2}$, 
common to {\it all} simply connected, conformally invariant, boundaries. 
This describes almost inaccessible sites: 
define ${\cal N}\left( H\right)$ as the 
number of boundary sites
having a given
probability $H$ to be hit by a RW starting at infinity;
the MF formalism yields,  
for
$H\rightarrow 0,$ a power law behavior
\begin{equation}
{\cal N}\left( H\right)|_{H\rightarrow 0}\approx H^{-(1+{n}^{\ast})}
\label{nha}
\end{equation}  
with an exponent 
\begin{equation}
1+n^{\ast}=\frac{23+c}{24}<1.
\end{equation}   

Some particular cases are worth considering. An {\it Ising} cluster possesses 
a multifractal spectrum with respect to the harmonic measure ($c=\frac{1}{2}$):
\begin{eqnarray}
\tau \left( n\right)&=&\frac{1}{2}\left( n-1\right) +\frac{7}{48}\left( 
\sqrt{48n+1}-7\right) , \label{taufis}\\
f\left( \alpha \right) &=&\frac{49}{96}\left( 3-\frac{1}{2\alpha -1}\right) -%
\frac{\alpha }{48},\quad \alpha \in \left( 
{\textstyle{1 \over 2}}%
,+\infty \right), \label{fis}
\end{eqnarray}
with a dimension of the accessible perimeter
\begin{equation}
D_{\rm EP}={\rm sup}_{\alpha}f(\alpha, c=\frac{1}{2})=\frac{11}{8}.
\end{equation}
The $Q=4$ Potts model provides an interesting example of a {\it left-handed} multifractal spectrum ($c=1$)
\begin{eqnarray}
\tau \left( n\right)&=&\frac{1}{2}\left( n-1\right) + 
\sqrt{n}-1, \label{taufc}\\
f\left( \alpha \right) &=&\frac{1}{2}\left( 3-\frac{1}{2\alpha -1}\right),\quad \alpha \in \left( 
{\textstyle{1 \over 2}}%
,+\infty \right), \label{fc}
\end{eqnarray}
with accessible sites forming a set of Hausdorff dimenson
\begin{equation}
D_{\rm EP}={\rm sup}_{\alpha}f(\alpha,c=1)=\frac{3}{2},
\end{equation}
which is also the {\it maximal} value common to all multifractal generalized dimensions 
$D(n)=\frac{1}{n-1}\tau(n)$. Notice that the external perimeter which bears 
the electrostatic charge is a {\it simple} curve, i.e. a curve without double points, 
a self-avoiding or {\it simple} path. We therefore arrive at the striking conclusion that in the plane, a
conformally invariant scaling curve which is self-avoiding has a Hausdorff dimension
at most equal to $D_{\rm EP}=3/2$\cite{duplantier11}. The  corresponding $Q=4$ Potts frontier, while still
possessing a set of double points of dimension $0$, actually develops
a logarithmically growing number of double points \cite{aharony}.\\

\textbf{\large{\textbf{ X Geometric Duality in $O(N)$ and Potts Cluster Frontiers}}}\\

The $O(N)$ model partition function
is that of a gas
$\cal G$
of
self- and mutually-avoiding {\it loops} on a given lattice, e.g.,
the hexagonal
lattice \cite{nien}: 
${Z}_{O(N)} = \sum_{\cal G }K^{{\cal N}_{B}}N^{{\cal N}_{P}},$ 
with $K$ and $N$ two fugacities, associated respectively with the 
total number of occupied bonds 
${\cal N}_{B}$, and with the total number 
${\cal N}_{P}$ of loops, 
i.e., polygons drawn on the lattice. 
For $N \in [-2,2]$, this model possesses a critical point (CP), $K_c$, 
while the whole {\it ``low-temperature''} (low-$T$) phase, i.e., ${K}_c < K$,
has critical universal properties, where the loops are {\it denser} 
that those at the critical point\cite{nien}. 
   
The partition function of the $Q$-state Potts model on, e.g., 
the square lattice, with a second order critical point for $Q \in [0,4]$, has a 
Fortuin-Kasteleyn representation {\it at} the CP: 
$ Z_{\rm Potts}=\sum_{\cup (\cal C)}Q^{\frac{1}{2}{\cal N}_{P}},$
where the configurations $\cup (\cal C)$ are those of reunions of 
clusters on 
the square lattice, with 
a total number ${\cal N}_{P}$ of polygons encircling all clusters, 
and filling the medial square lattice of the original lattice \cite{nien,dennijs}. 
Thus the critical Potts model becomes a {\it dense} loop model, with a loop fugacity 
$N=Q^{\frac{1}{2}}$, while one can show that its {\it tricritical} point with site 
dilution corresponds to the $O(N)$ CP\cite{D6}. 
The $O(N)$ and Potts models thus 
possess the same ``Coulomb gas'' 
representations \cite{nien,dennijs,D6}:
\begin{equation}
N=\sqrt{Q}=-2 \cos \pi g, 
\nonumber
\end{equation}
with $g \in [1,\frac{3}{2}]$ for the $O(N)$ CP, and $ g \in [\frac{1}{2},1]$ for the low-$T$ $O(N)$, 
or critical Potts, models;
the coupling constant $g$ of the Coulomb gas  
yields also the central charge:
\begin{equation}
c=1-6{(1-g)^2}/{g}. 
\label{cg}
\end{equation}

\noindent Notice that from the expression (\ref{cgamma}) of $c$ in terms of $\gamma \leq 0$ one 
arrives at the simple relation:
\begin{equation}
\gamma=1-g,\ g \geq 1;\ \gamma=1-1/g,\ g\leq 1.
\label{ggamma} 
\end{equation}
The above representation for $N=\sqrt Q \in [0,2]$ gives 
a range of values $- 2 \leq c \leq 1$; our results also apply for $c \in(-\infty, -2]$, 
corresponding, e.g., to the $O\left(N\in [-2,0]\right)$ branch, with a low-$T$ phase for $g \in [0,\frac{1}{2}]$, 
and the CP for $g \in [\frac{3}{2},2].$
 
The fractal dimension $D_{\rm EP}$ of the accessible perimeter, Eq. (\ref{D(c)}), is, once
 rewritten in terms of $g$, and like $c(g)=c(g^{-1})$, a symmetric function of $g$
\begin{equation}
D_{\rm EP}=1+ \frac{1}{2}g^{-1}\vartheta(1-g^{-1})+\frac{1}{2}g \vartheta(1-g),
\label{DEP}
\end{equation}
where $\vartheta$ is the Heaviside distribution, thus 
given by two different analytic
expressions on either side of the separatrix $g=1$. 
The dimension of the {\it hull's frontier}, i.e., the complete set of outer boundary 
sites of a cluster, has been 
determined for $O(N)$ and Potts clusters \cite{SD}, and reads 
\begin{equation}
D_{\rm H}=1+\frac{1}{2}g^{-1},
\label{DH}
\end{equation}
for the {\it entire} range of the coupling constant $g \in [\frac{1}{2},2]$. 
Comparing to Eq. (\ref{DEP}), 
we therefore see that the accessible perimeter and hull dimensions {\it coincide} for $g\ge 1$, i.e., 
at the $O(N)$ CP (or for tricritical Potts clusters), whereas they {\it differ}, namely $D_{\rm EP} <
D_H$, for $g < 1$, i.e., in the $O(N)$ low-$T$ phase, 
or for critical Potts clusters. 
This is the generalization to any Potts model of the effect originally found  
in percolation \cite{GA}. This can be directly understood in terms of the
{\it singly connecting} sites (or bonds) where fjords close in the scaling limit. 
Their dimension is given by\cite{SD}
\begin{equation} 
D_{\rm SC}=1+\frac{1}{2}g^{-1}-\frac{3}{2}g.
\label{DSC}
\end{equation}
Thus, for critical $O(N)$ loops, $g \in (1,2]$ and $D_{\rm SC} < 0,$ so there exist no closing fjords, 
which explains the 
identity:
\begin{equation}
D_{\rm EP} = D_{\rm H};
\label{id}
\end{equation} 
whereas $D_{\rm SC} > 0, g \in [\frac{1}{2},1)$ for critical Potts clusters, 
or in the $O(N)$ low-$T$ phase, 
where pinching points of positive dimension appear in the scaling limit, 
so that $D_{\rm EP} < D_{\rm H}$ (Table 1). we then find from Eq. (\ref{DEP}), with $g\leq 1$:
\begin{equation}
\left(D_{\rm EP}-1\right) \left( D_{\rm H}-1\right)=\frac{1}{4}.
\label{duali}
\end{equation}
The symmetry point $D_{\rm EP} = 
D_{\rm H}=\frac{3}{2}$ corresponds to $g=1$, $N=2$, or $Q=4$, 
where, as expected, the dimension $D_{\rm SC}=0$ of the pinching points 
vanishes.

For percolation, described either by $Q=1$, or by the low-$T$ $O(N=1)$ model, with 
$g=\frac{2}{3}$, we recover the result $D_{\rm EP}=\frac{4}{3}$, recently derived in \cite{DAA}. 
For the Ising model, described either by $Q=2, g=\frac{3}{4}$, or by 
the $O(N=1)$ CP, 
$g'=g^{-1}=\frac{4}{3}$, we observe 
that the EP dimension $D_{\rm EP}=\frac{11}{8}$ coincides, as expected, with that of the critical 
$O(N=1)$ 
loops, which in fact appear as EP's. This is a particular case of the geometric {\it duality} relation 
between the external perimeters of critical Potts models and the loops of $O(N)$ models at their critical points:
\begin{equation}
 D_{\rm EP}\left(Q(g)\right)=
D_{\rm H}\left(O\left(N(g')\right)\right), \, {\rm for}\, \, g'=g^{-1},\, g \le 1\ .
\end{equation}
In terms of this duality, the central charge takes the simple expression:
\begin{equation}
c=(3-2g)(3-2g').
\label{cdual}
\end{equation}

\begin{table}[t]
\begin{tabular}{| c | c | c | c | c | c |}
$Q$          &     0     &    1     &      2      &      3      &  4          \\
\hline
$c$          &     -2    &    0     &     1/2     &      4/5       &  1        \\
\hline
$D_{\rm EP}$ & ${5}/{4}$ &${4}/{3}$ & ${11}/{8}$  & ${17}/{12}$ & ${3}/{2}$ \\
\hline
$D_{\rm H}$  & $2$       &${7}/{4}$ & ${5}/{3}$   & ${8}/{5}$   & ${3}/{2}$ \\
\hline
$D_{\rm SC}$ & ${5}/{4}$ &${3}/{4}$ &$ {13}/{24}$ & ${7}/{20}$  &   $ 0$        \\
\end{tabular}
\medskip
\caption{Dimensions for the critical $Q$-state Potts model; $Q=0,1,2$ 
correspond respectively
to spanning trees, percolation and Ising clusters.}
\end{table}

\textbf{\large{\textbf{ XI Relation to the $SLE_{\kappa}$ process}}}\\

  An introduction to the stochastic L\"owner evolution process ($SLE_{\kappa}$) 
 can be found in \cite{lawleresi}. This process essentially describes the cluster {\it hulls} 
 we have introduced above. They can be simple or self-intersecting paths. 
 The $SLE_{\kappa}$ is parameterized by $\kappa$, which describes the rate of an auxiliary Brownian 
 motion, which is the source to the process. When $\kappa \in [0,4]$, the random curve is simple, 
 while for $\kappa \in (4,8)$, the curve is a self-intersecting path. 
 For $\kappa > 8$ the path is space filling.
 The correspondence to our parameters, 
 the central charge $c$, the string susceptibility exponent $\gamma$, or 
 the Coulomb gas constant $g$, is as follows.
 
 In the original work by Schramm \cite{schramm1}, the variance
  of the Gaussian winding angle of a $SLE_{\kappa}$ of size $R$ was calculated, 
 and found to be $\sqrt {\kappa {\rm ln} R}$. In \cite{DSw} we found, for instance for the $O(N)$ model, 
 the variance $\sqrt {(4/g) {\rm ln} R}$, from which we immediately infer the identity
 \begin{equation}
 \kappa=\frac{4}{g}\ .
 \label{k}
 \end{equation}
  
The low-temperature branch $g \in [\frac{1}{2},1)$ of the $O(N)$ model, for $N\in [0,2)$, indeed corresponds 
to $\kappa \in (4,8]$ and describes non simple curves, while $N\in [-2,0], g\in [0,\frac{1}{2}]$ 
corresponds to $\kappa \geq 8$. The critical point branch $g \in [1,\frac{3}{2}], N\in [0,2]$
 gives $\kappa \in [\frac{8}{3},4]$, while $g \in [\frac{3}{2},2], N\in [-2,0]$
 gives $\kappa \in [2,\frac{8}{3}]$.
The range $\kappa \in [0,2)$ probably corresponds to higher multicritical points with $g>2$.
Owing to Eq. (\ref{ggamma}) for $\gamma$, we have
  
\begin{equation}
\gamma=1-\frac{4}{\kappa},\ \kappa \leq 4\ ;\ \gamma=1-\frac{\kappa}{4},\ \kappa \geq 4\ .
\end{equation}
 The central charge (\ref{cgamma}) or (\ref{cg}) reads accordingly:
\begin{equation}
c=1-24{\left(\frac{\kappa}{4}-1\right)^2}/{\kappa}\ ,
\label{ck}
\end{equation}
an expression which of course is symmetric under the {\it duality} $\kappa/4 \to 4/\kappa=\kappa'$, or 
$$\kappa \kappa'=16\ ,$$ reflecting the symmetry under $gg'=1$ \cite{duplantier11}.  
The self-dual form of the central charge reads accordingly:
\begin{equation}
c=\frac{1}{4}(6-\kappa)(6-\kappa').
\label{cdualSLE}
\end{equation}
  
From Eqs. (\ref{DH}) and (\ref{DEP}) we respectively find
\begin{equation}
D_{\rm H}=1+\frac{1}{8}\kappa\ , 
\label{DHs}
\end{equation} 
\begin{equation}
D_{\rm EP}=1+ \frac{2}{\kappa}\vartheta(\kappa-4)+\frac{\kappa}{8} \vartheta(4-\kappa)\ ,
\label{DEPs}
\end{equation}
in agreement with later derivations in probability theory \cite{schramm2}. For $\kappa \leq 4$, we have 
$D_{\rm EP}(\kappa)=D_{\rm H}(\kappa)$. For $\kappa \geq 4$, the 
self-intersecting scaling paths obey the duality equation (\ref{duali}) 
derived above, recast here in the context of the $SLE_{\kappa}$ process:
\begin{equation}
\left[D_{\rm EP}(\kappa)-1\right] \left[ D_{\rm H}(\kappa)-1\right]=\frac{1}{4},\ \kappa \geq 4\ ,
\label{dualibis}
\end{equation}
where now $$D_{\rm EP}(\kappa)=D_{\rm H}(\kappa'=16/\kappa)\quad \kappa'\leq 4\ .$$ Thus 
we predict that the external perimeter of a self-intersecting $SLE_{\kappa \geq 4}$ process is by {\it duality} 
the simple path of the $SLE_{(16/{\kappa})=\kappa' \leq 4}$ process.

The symmetric point $\kappa=4$ corresponds to the $O(N=2)$ model, or $Q=4$ Potts model, with $c=1$. 
The value $\kappa=8/3, c=0$ corresponds to a self-avoiding walk, 
which thus appears \cite{duplantier9,DAA} as the external frontier of a $\kappa=6$ process,
namely that of a percolation hull \cite{schramm1,smirnov1}.
 
 Work remains to be done to elucidate the relation between the $SLE$ 
 construction in probability theory and the
  Coulomb gas and conformal invariance approaches, as well as the quantum gravity method 
  described here.

\end{document}